\documentclass[journal,transmag]{IEEEtran}
\ifCLASSINFOpdf
\else
\fi
\usepackage{cite}
\usepackage{amsmath,amssymb,amsfonts}
\usepackage{algorithmic}
\usepackage{graphicx}
\usepackage{textcomp}
\usepackage{xcolor}
\usepackage{mhchem}
\usepackage{siunitx} 
\usepackage{balance}
\usepackage{hyperref}
\usepackage{subfigure}
\usepackage{array}
\usepackage{booktabs}
\usepackage{multirow}
\usepackage{colortbl}
\usepackage{tabularx}
\definecolor{lightgray}{gray}{0.91}
\usepackage{setspace}
\usepackage{graphicx}
\def\BibTeX{{\rm B\kern-.05em{\sc i\kern-.025em b}\kern-.08em
		T\kern-.1667em\lower.7ex\hbox{E}\kern-.125emX}}
\usepackage{tikz}
\usetikzlibrary{arrows}
\tikzstyle{int}=[draw, fill=white!20, minimum width=3cm, minimum height=1cm]

\begin{document}
	\def\endthebibliography{%
		\def\@noitemerr{\@latex@warning{Empty `thebibliography' environment}}%
		\endlist
	}
\sloppy

\title{Spheroidal Molecular Communication via Diffusion: Signaling Between Homogeneous Cell Aggregates}

\author{\IEEEauthorblockN{Mitra Rezaei, Hamidreza Arjmandi, Mohammad Zoofaghari, \\
		Kajsa Kanebratt, Liisa Vilén, David Janzen, Peter Gennemark, and Adam Noel, \IEEEmembership{Member,~IEEE}}

\thanks{This work was supported by the Engineering and Physical Sciences Research Council [grant number EP/V030493/1]. For the purpose of open access, the authors have applied a Creative Commons Attribution (CC-BY) licence to any Author Accepted Manuscript version arising from this submission.} 

\thanks{M.~Rezaei and A.~Noel are with the School of Engineering, University of Warwick, Coventry, UK. (e-mail: \{Mitra.Rezaei, Adam.Noel\} @warwick.ac.uk.)}

\thanks{ H.~Arjmandi is with the Institute of Cancer and Genomic Sciences, University of Birmingham, Birmingham, UK. e-mail: h.arjmandi@bham.ac.uk.}

\thanks{ M.~Zoofaghari is with the Electrical Engineering Department, Yazd University, Yazd, Iran. e-mail: zoofaghari@yazd.ac.ir.}
	
\thanks{K.~Kanebratt and L.~Vilén are with the Drug Metabolism and Pharmacokinetics, Research and Early Development, Cardiovascular, Renal and Metabolism (CVRM), Bio Pharmaceuticals, R\&D, AstraZeneca, Gothenburg, Sweden. (e-mail: \{Kajsa.P.Kanebratt, liisa.vilen @astrazeneca.com.\})}

\thanks{D.~Janzen is with the Clinical Pharmacology and Quantitative Pharmacology, Clinical Pharmacology and Safety Sciences, AstraZeneca AB R\&D Gothenburg, Gothenburg, Sweden. e-mail: david.janzen@astrazeneca.com.}
 
\thanks{P.~Gennemark is with the Drug Metabolism and  Pharmacokinetics, Research and  Early Development, Cardiovascular, Renal and Metabolism (CVRM), Bio-Pharmaceuticals, R\&D, AstraZeneca, Gothenburg, Sweden and the Department of Biomedical Engineering, Link\"oping University, Sweden. e-mail: Peter.Gennemark@astrazeneca.com.}
}


\maketitle
\begin{abstract}
Recent molecular communication (MC) research has integrated more detailed computational models to capture the dynamics of practical biophysical systems. This paper focuses on developing realistic models for MC transceivers inspired by spheroids – three-dimensional cell aggregates commonly used in organ-on-chip experimental systems. Potential applications that can be used or modeled with spheroids include nutrient transport in organ-on-chip systems, the release of biomarkers or reception of drug molecules by cancerous tumor sites, or transceiver nanomachines participating in information exchange.
In this paper, a simple diffusive MC system is considered where a spheroidal transmitter and spheroidal receiver are in an unbounded fluid environment. These \textit{spheroidal antennas} are modeled as porous media for diffusive signaling molecules, then their boundary conditions and effective diffusion coefficients are characterized. Furthermore, for either a point source or spheroidal transmitter, the Green's function for concentration (GFC) outside and inside the receiving spheroid is analytically derived and formulated in terms of an infinite series and confirmed with a particle-based simulator (PBS). The provided GFCs enable computation of the transmitted and received signals in the proposed spheroidal communication system. This study shows that the porous structure of the receiving spheroid amplifies diffusion signals but also disperses them, thus there is a trade-off between porosity and information transmission rate. 
Furthermore, the results reveal that the porous arrangement of the transmitting spheroid not only disperses the received signal but also attenuates it in comparison to a point source transmitter. System performance is also evaluated in terms of the bit error rate (BER).
Decreasing the porosity of the receiving spheroid is shown to enhance the system performance. Conversely, reducing the porosity of the transmitting spheroid can adversely affect system performance.
\end{abstract}

\begin{IEEEkeywords}
Molecular communication, Spheroid, Organ-on-chip, Diffusion, Spheroidal communication, Realistic transceiver.
\end{IEEEkeywords}

\IEEEdisplaynontitleabstractindextext
\section{Introduction}
Molecular communication (MC) is a bio-inspired mechanism that is envisioned to realize micro- and nano-scale communication systems using molecules as information carriers \cite{farsad2016comprehensive}. 
Despite many efforts by the MC community to model the components of MC systems, more realistic models are required. Existing literature over-simplifies MC components in biological environments and does not sufficiently account for how signaling molecules interact with biological or biosynthetic transmitters and receivers (i.e., cells) or with the environment. Elements and structures of \textit{in vitro} environments such as Organs-on-Chip (OoCs) could be used to improve model realism, to provide mechanistic insight into the OoC biology, and more generally to contribute to MC research and development. For example, the transmitter could be a group of beta-cells emitting insulin and the receiver could be a group of liver cells that detect the insulin signal and react by increasing its glucose uptake. 

Previous MC works have theoretically modeled and simulated the transmitter as a single cell (or machine) releasing molecules with different release mechanisms and the receiver as a single cell detecting the molecules. 
However, cells do not normally live in isolation but in diverse populations with other cells. This is true \textit{in vivo} but also in many \textit{in vitro} systems. In particular, tissues and tumors in multi-cellular organisms and biofilms of microorganisms are common natural instances whereas spheroids, organoids, tumoroids, and islet microtissues are well-known instances in biological experimental setups. Thus, we are inspired by the design of MC transceivers based on a population of (biological or biosynthetic) cells.
One realistic transceiver for MC is a spheroid structure, which is a 3D cell aggregation of hundreds or thousands of cells in a spherical shape that is widely used in Organ-on-Chip systems, e.g., in \cite{bauer2017functional}. 
These \textit{microphysiological systems} have the promising capability to emulate natural organ-to-organ communication dynamics \textit{in vitro}. For example, a liver-pancreas OoC model with organ-to-organ communication was demonstrated in \cite{bauer2017functional}, with the purpose of providing a distinctive platform for studying type 2 diabetes mellitus. In this system, glucose is controlled through communication between liver spheroids and pancreatic islets. The pancreatic islets release insulin under high glucose conditions, which increases glucose uptake by the liver spheroids and reduces their glucose release.
The liver-pancreas OoC inspires biosynthetic MC systems for healthcare applications. As a first attempt to realize this spheroid-based MC system and study its communication dynamics, we require a suitable biophysical model for an individual spheroid.

Several studies have attempted to model the process of mass transport in a spheroidal environment. 
The authors of \cite{astrauskas2019modeling} modeled the penetration and diffusion of dyes within multicellular spheroids using diffusion-reaction equations. They assumed constant concentrations of molecules at the inside and outside of the outer spheroid's boundary. Paper \cite{leedale2020multiscale} focused on mathematically modeling the spatiotemporal dynamics of drugs in spheroids. The authors investigated how drug characteristics impact the propagation of the drug inside a spheroid. 
As boundary conditions at the spheroid border, they assumed flow continuity and that the concentrations on either side of the outer spherical boundary were equal. 
The authors of \cite{bull2020mathematical} developed an agent-based mathematical model to simulate the passive distribution of microbeads into tumor spheroids while considering the spheroid's growth through externally-supplied oxygen. In their approach, they assumed a constant oxygen concentration at both sides of the spheroid's outer boundary. 
Paper \cite{graff2003theoretical} proposed a model to investigate how factors including the association and dissociation rates, degradation rate, and plasma clearance rate influence the depth of antibody penetration in tumor spheroids. As a boundary condition, they considered the concentration of unbound antibodies inside the spheroid to be equal to that outside multiplied by the porosity parameter or the fraction of tumor volume accessible to the antibody.
The authors of \cite{goodman2008spatio} developed a mathematical model to examine the diffusion of nanoparticles into tumor spheroids that considers the structural nonuniformity of the spheroid in the radial direction. They applied a boundary condition that determines the concentration inside the spheroid by multiplying the concentration outside with the radially-dependent spheroid porosity. In \cite{grimes2014method}, the researchers presented a mathematical model to predict oxygen diffusion and consumption in tumor spheroids. They employed an analytical solution based on the spherical reaction-diffusion equation with a constant concentration at the tumor boundary that is not dependent on the surrounding concentration.

Furthermore, in our recent study in \cite{arjmandi20233d}, we mathematically proved that the amplification is indeed present at the boundary and derived the corresponding amplification factor. The provided theorem was investigated through experiments using liver spheroids and glucose as the target molecules. Interestingly, a rapid decrease in the concentration of molecules in the surrounding medium was observed when a spheroid was introduced to a culture medium with a higher glucose concentration. 
The distinct propagation characteristics prompt us to demonstrate whether spheroid properties facilitate communication. In this direction, we next review relevant existing literature on modeling MC transceiver components.

\textit{Transmitters:} The most common transmitter model is an ideal point source that releases molecules instantaneously, disregarding physical geometry and realistic release mechanisms \cite{jamali2019channel}. Paper \cite{garralda2011simulation} pioneered the concept of a pulse-shaped release from a point source transmitter, in which multiple molecules can be released during the pulse. Alternatively, the volume transmitter model proposed in \cite{noel2016channel} incorporated the transmitter's geometric shape but did not consider a distinct transmitter boundary, leading to an equal concentration inside and outside the transmitter's boundary. Paper \cite{pierobon2010physical} studied a box-like transmitter with a surface outlet for controlling molecule release to establish desired concentration gradients. Furthermore, \cite{chude2015diffusion} assumed a spherical structure with nanopores for molecule passage. The model in \cite{arjmandi2016ion} incorporated ion channels on a cell surface transmitter sensitive to either electrical voltage or ligand concentration variations for molecule release. Storage and production dynamics inside the transmitter have also been explored in the literature, ranging from instantaneous \cite{bafghi2018diffusion} to pulse function release \cite{khaloopour2019adaptive} and (inspired by neuronal behavior) exponential release based on the number of molecules stored, as proposed in \cite{rezaei2022molecular}. Paper \cite{Huang2022} introduced a transmitter model that employs membrane fusion. This mechanism involves the fusion of a vesicle, loaded with molecules and formed within the transmitter, with the transmitter membrane, which enables molecules to be released into the extracellular environment. Moreover, the authors of \cite{MAX2022} investigated the controlled release of signaling molecules from compact, non-porous matrix-type drug carriers. The authors also studied the gradual release of molecules from the matrix structure, considering a perfect sink condition at the matrix boundary. However, one can extend their study to porous matrices by incorporating modified diffusion coefficients and drug solubility \cite{FRENNING201188, YIN201178, higuchi1963mechanism}.

\textit{Receivers:} The passive receiver, in which information molecules freely diffuse in the receiver's space and the movement of molecules is not affected, is the most common model used in previous works \cite{jamali2019channel}. Such a simple model is often used to facilitate analysis of other aspects of an MC system, e.g., the environment boundary \cite{arjmandi2019diffusive}. 
The passive model may be relevant for small and hydrophobic molecules (which are repelled from water molecules) that easily pass through a cell membrane. 
However, many extracellular molecules are too large or too lipophobic to traverse the cell membrane \cite{bi2021survey} and also cells usually react with signaling molecules (either directly on the surface or in the intracellular environment). 

To address the limitations of the passive receiver model, some works have considered a reception mechanism across the receiver (cell) membrane to activate an internal signaling pathway (i.e., a series of chemical reactions controlling a cell function). These works, including \cite{sabu20203,pierobon2011noise,akkaya2014effect,chou2015impact,lotter2021saturating,genc2018reception}, have studied the effects of various reaction mechanisms across the membrane, including cell membrane receptors that can vary in size, number, and spatial distribution. 
 
MC system analysis for microfluidic systems has not investigated relatively large transceivers with different diffusion coefficients. This paper is the initial work in the field to consider spheroidal transceivers under corresponding boundary conditions.
However, to establish a complete microfluidic system with spheroids as transceivers, we need to characterize the diffusion and convection of molecules through microfluidic channels and wells, as well as their diffusion and convection within the porous spheroidal structures. In this study, we are primarily concerned with a simplified version of the problem at hand. Hence, in this work, our focus is to model communication between two spheroidal structures, inspired by the cross-talk phenomenon studied in \cite{bauer2017functional}. We consider this communication in an unbounded environment without fluid flow. Nevertheless, this assumption is reasonable because the dimensions of the spheroids and the distance between them are significantly smaller than the size of the well in a typical microfluidic system (e.g., 5\,mm diameter according to \cite{schimek2013integrating}). Moreover, the average flow rate is extremely low in a reference study on which we based our model (i.e., $4.94\,\mu \rm L/\rm min$ with a complete media turnover time of approximately $2\, \rm h$ according to \cite{bauer2017functional}), supporting our decision to not include convection in this initial work. This research highlights how the distinct features of spheroid structure support the potential application of \textit{spheroidal antennas} in synthetic biology for transmitting and receiving information.
The spheroids' ability to release molecules in a controlled manner enables them to function as transmitters. The porous spheroidal structure also increases the received power of the diffusion signal. Use cases can include improved communication between drug carriers formed from porous matrices for precisely locating diseased cells.  

In this paper, we introduce an end-to-end diffusive MC system using spheroids as both transmitter and receiver. The contributions of this paper are summarized as follows: 
\begin{itemize}
\item We propose a novel spheroid-to-spheroid (S2S) communication system in an unbounded environment.
\item We formulate the molecule release function from the border of a transmitting multicell spheroid as a boundary value problem and derive the corresponding Green's function for concentration (GFC). 
\item As we did for the impulsive point source transmitter in \cite{arjmandi2023diffusive}, the joint communication channel and spheroidal receiver's response are also modeled as a boundary value problem that accounts for the amplification at the boundary.
\item We validate our results with particle-based simulations (PBS), confirming the accuracy and reliability of our model.
\item We study the geometric effects of the transmitting and receiving spheroids on the molecule concentration inside the receiving spheroid.
\item We evaluate the system's bit error rate (BER) in the presence of inter-symbol interference (ISI) with different time slot durations and spheroid porosity levels.
\end{itemize}

The rest of the paper is organized as follows. Section \ref{SM} describes the spheroid structure and diffusive MC system using the spheroid. In Section \ref{GFB}, the spheroid Green's function is provided. The details regarding the characterization of the diffusive MC channel are presented in Section \ref{S2S}. Results and discussions are outlined in Section \ref{results}. Finally, Section \ref{Conclusion} concludes the paper.

\section{System Model}\label{SM}
\subsection{Diffusive MC System with Spheroidal Transceivers}
\begin{figure}[t]
	\centering
	\includegraphics[width=3.5in]{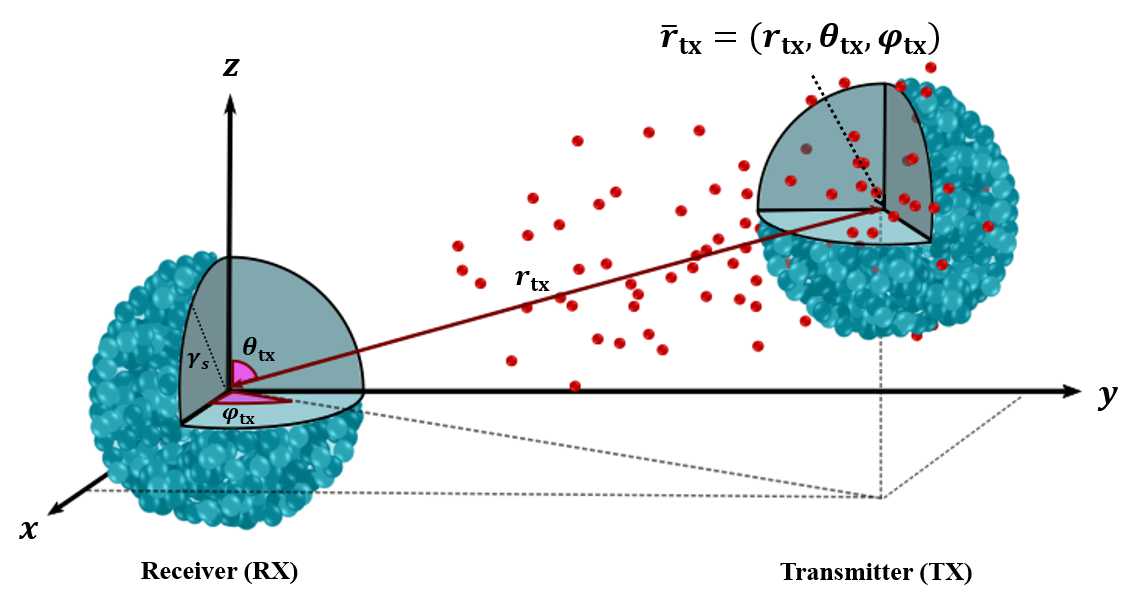}
	\caption{System model schematic where the red and greenish-black spheres represent the signaling molecules and cells, respectively. The receiving spheroid is centrally positioned at the coordinate system's origin, and the transmitting spheroid is centered at a distance of $r_{\rm tx}$ along the coordinate system.}
	\label{Fig0}
\end{figure}

An end-to-end diffusive MC system is considered, where both the transmitter and receiver are spheroids. We consider two spheroids at fixed locations within an unbounded fluid medium, as shown in Fig. \ref{Fig0}. The spherical coordinate system, characterized by $(r,\theta,\varphi)$, is used to illustrate the geometry of the environment, where $r$, $\theta$, and $\varphi$ represent the radial, elevation, and azimuthal coordinates, respectively. The origin of the spherical coordinate system is set at the center of the receiving spheroid, while the transmitting spheroid is centered at the arbitrary point $\bar{r}_{\rm tx}=(r_{\rm tx},\theta_{\rm tx},\varphi_{\rm tx}).$ The transmitter uses signaling molecules of type \textit{A}. 
The released molecules move randomly within the environment of the spheroidal transmitter by following Brownian motion before eventually diffusing into the surrounding medium where they continue to freely diffuse. The molecule movements are assumed to be independent of each other. Some of the diffusing molecules will pass into (and potentially out of) the receiving spheroid.

The molecules may bind to cell receptors through a chain of reactions leading to product molecules $E$ that are detected by cells. We do not consider this mechanism in the model and approximate the process based on first-order degradation, i.e.,
\begin{align}\label{RC1}
\ce{$A$ -> $E$}.
\end{align}

We focus on investigating the concentration profile of type $A$ molecules outside and within the spheroid.

In practice, spheroids are constructed by various methods with the aim to emulate the internal physiological activities of an organ \cite{alberts2015essential}. This compact spheroidal structure forms by a spontaneous aggregation of cells where the integrins on their surfaces bind to the extracellular matrix (ECM) of other nearby cells \cite{gunti2021organoid}. When spheroids reach maturity, they are collected and placed into designated wells. If they grow too large, then oxygen diffusion to the core becomes restricted and this can lead to cell death and the formation of a necrotic center \cite{nath2016three}. However, in this research, our focus is on small, mature spheroids with a dense arrangement of cells.
This means that a signaling molecule moving within the extracellular space (ECS) would always be very close to the cells' boundaries and frequently exposed to bind to receptors. The ECS is the space between cells and is for exchanging nutrients, gases, and signaling molecules between the cells and the surrounding environment. Here, we assume that molecules within the ECS of the spheroid are homogeneously available for transmembrane reactions. We also note that this study does not incorporate the growth and death of cells within the spheroid due to the considerably lower rates of cell growth and death compared to the information transmission rate under consideration \cite{greenspan1972models}.
\subsection{Spheroidal Model}
As shown in Fig.~\ref{Fig0}, we consider a spheroid as a three-dimensional arrangement of $N_\mathrm{c}$ cells in a sphere-like shape with radius $\gamma_\mathrm{s}$ that mimics the structural and functional characteristics observed in the biological organs or tissue constructs that are used in laboratory environments. The interior space of the spheroid is composed of both the cells and the ECS. The ECS allows for the diffusion of molecules and the formation of concentration gradients. In this paper, we assume that the ECM materials do not contribute a meaningful volume to the occupancy of the spheroid and, consequently, simplify the ECS as empty fluid. Given that $V_\mathrm{c}$ represents the volume of an individual cell within the spheroid, the volumes of the cell matrix and the extracellular space between cells are equivalent to $V_\mathrm{c}N_\mathrm{c}$ and $V_\mathrm{s}-V_\mathrm{c}N_\mathrm{c}$, respectively, where the overall volume of the spheroid is $V_\mathrm{s}=\frac{4}{3}\pi \gamma_\mathrm{s}^3$.
In this study, we characterize the architecture of the spheroid as a porous medium, wherein the \textit{porosity parameter}, $\varepsilon$, serves as the ratio of the extracellular space to the overall volume of the spheroid, i.e.,  
\begin{align}\label{eps}
\varepsilon=1-\frac{N_\mathrm{c}V_\mathrm{c}}{V_\mathrm{s}}.
\end{align}

We assume that the spheroids are immersed in a fluid medium that occupies their extracellular space. Signaling molecules, nutrient molecules, and waste products diffuse within the ECS via curved paths, resulting in a shorter net molecule displacement over a given time interval than outside the spheroid. Consequently, the effective diffusion within the whole spheroid volume is reduced compared to the free fluid diffusion outside. We consider the entire spheroid volume to be a homogenized diffusion environment that we will refer to as the ``homogenized spheroid'' or simply ``spheroid'' throughout the remainder of this paper.
 The effective diffusion coefficient within this homogenized spheroid can be determined in terms of the diffusion coefficient ($D$) of the signaling $A$ molecules in the free fluid as \cite{FRENNING201188}
\begin{align}
D_{\rm eff}=\frac{\varepsilon}{\tau}D,
\end{align}
where $\tau$ is tortuosity, which refers to the degree of path irregularity or curvature experienced by a molecule while it traverses through the extracellular space of the spheroid, and is modeled as a function of spheroid porosity, i.e., $\tau=\frac{1}{\varepsilon^{0.5}}$ \cite{sharifi2019numerical}.

At the interface between the two diffusive environments with different diffusion coefficients, a continuity condition for flow must be satisfied, which is expressed as 
\begin{align}\label{BC2}
D_{\rm eff} \frac{\partial c_\mathrm{s}(\bar r,t)}{\partial r} =D \frac{\partial c_\mathrm{o}(\bar r,t)}{\partial r},
\end{align}
and another boundary condition that is generally modeled as \cite[Ch. 3]{crank1979mathematics}
\begin{align}\label{BC1}
c_\mathrm{s}(\bar r,t)=\kappa c_\mathrm{o}(\bar r,t),
\end{align}
where $ \bar r \in \partial \Omega$, $\partial \Omega$ denotes the spheroid boundary region, $\Omega$ is the spheroid region, and $c_\mathrm{s}$ and $c_\mathrm{o}$ denote the concentration function inside and outside the spheroid, respectively\footnote{We note that molecules can cross in both directions.}. The constant $\kappa$ is a function of the porosity of the medium and should be determined experimentally. In our results in Section \ref{results}, we demonstrate using a PBS that for two ideal diffusion environments with diffusion coefficients $D$ and $D_{\rm eff}$, we have $\kappa=\sqrt{\frac{D}{D_{\rm eff}}}$, which was theoretically verified and experimentally studied in \cite{arjmandi20233d}. Thus, for $\kappa\neq 1$, a concentration discontinuity (i.e., jump) occurs at the boundary.
 
 \begin{figure}[!t]
	\centering
	\includegraphics[width=4.3in]{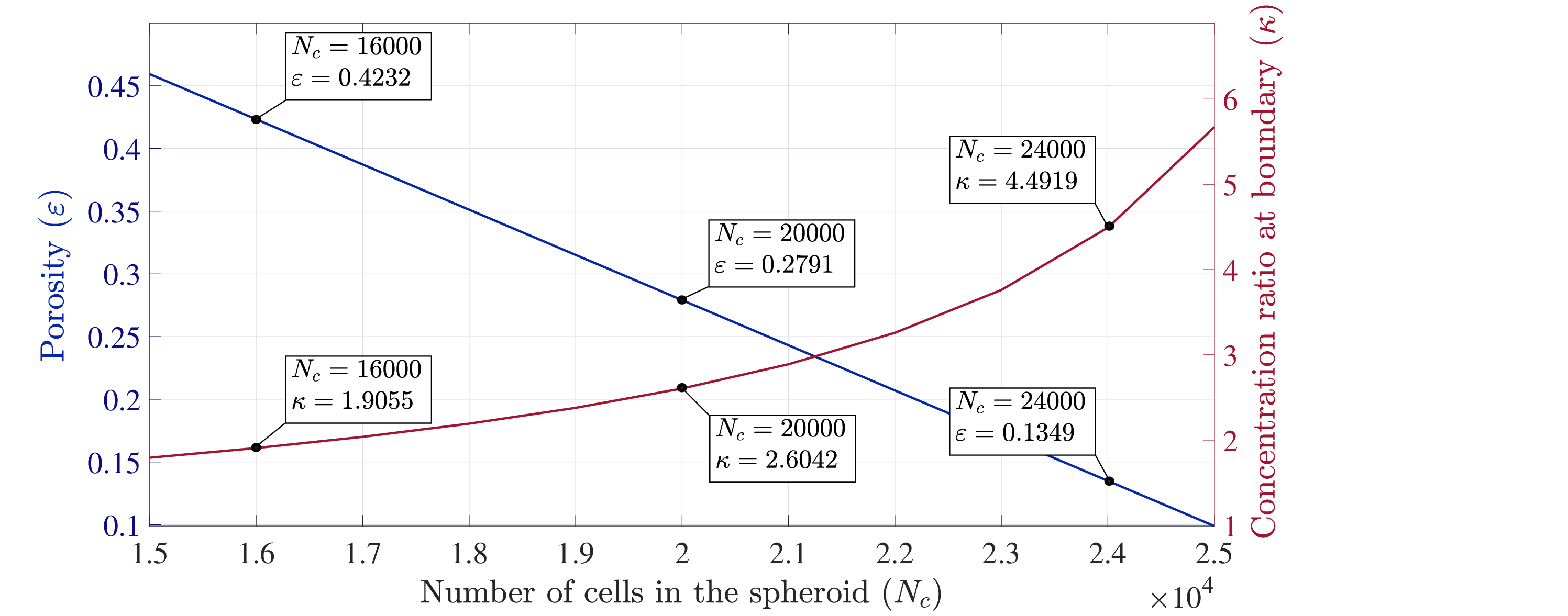}
	\caption{Porosity ($\varepsilon$) and concentration ratio at the boundary ($\kappa$) versus the number of cells $N_\mathrm{c}$ of fixed volume $V_\mathrm{c}=3.14\times 10^{-15}$\,\si{m^3} inside a spheroid of fixed radius $\gamma_\mathrm{s}=275$\,\si{\mu m}.}
	\label{fig1}
\end{figure}
Fig. \ref{fig1} demonstrates the porosity ($\varepsilon$) and boundary concentration ratio ($\kappa$) as a function of the number of cells in the spheroid, $15000<N_\mathrm{c}<25000$, when the spheroid radius and cell volume are assumed to be $R_\mathrm{s}=275$\,\si{\mu m} and $V_\mathrm{c}=3.14 \times 10^{-15}$\,\si{m^3}, respectively \cite{bauer2017functional}. As observed in Fig. \ref{fig1}, the boundary concentration ratio increases exponentially with an increase in $N_\mathrm{c}$. For $N_\mathrm{c}=24000$, which is the approximate number of cells in HepaRG spheroids \cite{bauer2017functional}, we have $\varepsilon=0.1349$ and correspondingly $\kappa=4.4919$. This value of $\kappa$ suggests a large concentration discontinuity at the spheroid boundary.

\section{GREEN’S FUNCTION FOR THE BOUNDARY VALUE DIFFUSION PROBLEM}\label{GFB}

To analyze the diffusive spheroidal MC system defined in Section~\ref{SM}, we divide the problem into two distinct parts: (i) modeling the transmitted signal given the release of molecules by the spheroidal transmitter with radius $\gamma_{\rm tx}$ and the effective diffusion coefficient of $D_{\rm eff}^{\rm tx}$; and (ii) modeling the received signal at the spheroidal receiver with radius $\gamma_{\rm rx}$ and the effective diffusion coefficient of $D_{\rm eff}^{\rm rx}$ given the transmitted signal.  
The Green's function for the boundary value diffusion problem is formulated for each section to characterize the \textit{expected} diffusion signal. Subsequently, these individual problems are combined into a single problem to analyze the end-to-end diffusive MC system.

\subsection{Diffusion Signal Generated by Spheroidal Transmitter}
The spheroidal transmitter is comprised of an aggregation of $N_{c}^{\rm tx}$ cells that are randomly distributed throughout the spheroidal environment to form a homogeneous porous medium. We assume that all cells release molecules simultaneously upon exposure to an external stimulus (e.g., light or sudden change in pH-level). 

We begin our analysis by formulating the transmitting spheroid's molecule release rate into an \textit{empty} environment, i.e., without the receiving spheroid. To simplify the characterization of the diffusion signal from the transmitting spheroid, we assume that the center of the transmitter corresponds to the origin of our coordinate system. First, we obtain the concentration of molecules due to instantaneous release from a point source transmitter positioned at an arbitrary location $\bar{r}_{0}=(r_0,\theta_0,\varphi_0)$ within the transmitting spheroid where one molecule is released at time $t_0$. This impulsive point source is represented by the function $S(\bar r,t,{\bar{r}_0},t_0)=\frac{{\delta (r  - {r_0})\delta(\theta-\theta_0)\delta(\varphi-\varphi_0)\delta(t-t_0)}}{ r^2\sin\theta  }$ $\si{s^{-1}.m^{-3}}$. Given the source $S(\bar r,t,{\bar{r}_0},t_0)$, the molecule diffusion \textit{inside the transmitting spheroid} is described by the partial differential equation (PDE) \cite{grindrod1996theory}
\begin{align}\label{tx_fick}
D_{\rm eff}^{\rm tx}{\nabla ^2}c_{\rm tx}(\bar r,t|\bar{r}_0,{t_0})
+ S(\bar r,t,\bar{r}_0,t_0) = \frac{\partial c_{\rm tx}(\bar r,t|{\bar{r}_0},{t_0})}{{\partial t}},
\end{align} 
where $c_{\rm tx}(\bar r,t|{\bar{r}_0},{t_0})$ denotes the molecule concentration at point $\bar r$ inside the transmitting spheroid at time $t$, given the aforementioned impulsive point source. In the spherical coordinate system, the Fourier transform of \eqref{tx_fick} is obtained as

\small
\begin{align}\label{tx_fick_fourier}
&\frac{D_{\rm eff}^{\rm tx}}{{{r^2}}}\frac{\partial }{{\partial r}}\left({r^2}\frac{{\partial C_{\rm tx}}}{{\partial r}}\right) +
\frac{D_{\rm eff}^{\rm tx}}{{{r^2}\sin \theta }}\frac{\partial }{{\partial \theta }}\left(\sin \theta \frac{{\partial C_{\rm tx}}}{{\partial \theta }}\right)+ \frac{D_{\rm eff}^{\rm tx}}{{{r^2}{{\sin }^2}\theta }}\frac{{{\partial ^2}C_{\rm tx}}}{{\partial {\varphi ^2}}}\\\nonumber
&+\frac{{\delta (r  - {r_0})\delta(\theta-\theta_0)\delta(\varphi-\varphi_0)\mathrm{e}^{-i\omega t_0}}}{ r^2\sin\theta  } = i\omega C_{\rm tx}, \nonumber
\end{align}
\normalsize
where $C_{\rm tx}$ is the Fourier transform of $c_{\rm tx}$, $i$ is the imaginary unit, and $\omega$ is the angular frequency variable in the Fourier transform\footnote{We use lowercase $c$ and uppercase $C$ to represent the concentration function in the time and frequency domains, respectively.}.
The free-space diffusion \textit{outside the transmitting spheroid} is modeled as 
\begin{equation}\label{out_fick_fourier}
D{\nabla ^2}C_{\rm otx}(\bar r,\omega|{\bar{r}_0},{t_0})= i\omega C_{\rm otx}(\bar r,\omega|{\bar{r}_0},{t_0}).
\end{equation}

In this equation, $C_{\rm otx}$ is the Fourier transform of $c_{\rm otx}$, which is the molecule concentration at point $\bar r$ outside the transmitting spheroid at time $t$. The concentration functions $c_{\rm tx}(\bar r,t|{\bar{r}_0},{t_0})$ and $c_{\rm otx}(\bar r,t|{\bar{r}_0},{t_0})$ that satisfy \eqref{tx_fick_fourier} and \eqref{out_fick_fourier} subject to the boundary conditions \eqref{BC2} and \eqref{BC1} are called the Green's functions for concentration (GFCs) of diffusion inside the transmitting spheroid and in the medium, respectively. We have solved the problem and obtained the series-form expression \eqref{Eqsd11} in Appendix \ref{APXA}.
Due to the linearity of the PDEs, we account for the molecule concentration resulting from \textit{all} transmitting cells by calculating the integral of $c_{q}$, $q\in\{\rm tx,\rm otx\}$, across the transmitter volume $\Omega_{\rm TX}$ as
\begin{equation}
c_q^{\rm Total}(\bar r,t)=\frac{1}{v_{\rm TX}}\int_{v^{'}\in \Omega_{\rm TX}} c_q(\bar r,t|{\bar{r}_0},{t_0}) dv^{'},
\end{equation}
where $\frac{1}{v_{\rm TX}}$ is for normalization. This numerical integral is evaluated with a discrete approximation that improves by increasing the resolution of the discretization.
The total summation of molecules inside ($N_{\rm in}(t)$) and outside ($N_{\rm out}(t)$) the transmitting spheroid is constant, represented as $N_\mathrm{c}^{\rm tx} \times N$, where $N$ is the number of molecules released per cell. By normalizing to $N_\mathrm{c}^{\rm tx} \times N$, we have $N_{\rm in}(t) + N_{\rm out}(t)=1$. 
The release rate is defined as either the negative of the change in the number of molecules inside the transmitter $\left(-\frac{dN_{\rm in}(t)}{dt}\right)$ or the change in the number of molecules outside the transmitter $\left(\frac{dN_{\rm out}(t)}{dt}\right)$. $N_{\rm out}$ is obtained by integrating the concentration of total molecules outside the spheroid. Consequently, the \textit{expected} release rate $g(t)$ is computed as 
\begin{equation}\label{rlsrate}
g(t)=\frac{d}{dt} \left( \int_{\theta=0}^{\pi} \int_{\phi=0}^{2\pi} \int_{r=\gamma_{\rm tx}}^\infty c_{\rm otx}^{\rm Total} r^2 \sin\theta dr d\phi d\theta \right).
\end{equation}

For tractability in the receiver analysis, we assume that the distance separating the transmitter and receiver is much larger than the sizes of the spheroids themselves. This approximation is consistent with the findings in \cite{noel2016channel}, where the authors compare a spherical transmitter to a point source transmitter and illustrate the improved approximation at greater distances. With this approximation, we can assume that the transmitting spheroid can be represented as a point source with the release function $g(t)$. However, to investigate signal propagation between two nearby spheroids, it becomes essential to consider a volume transmitter with release function $g(t)$ at the spheroid's surface. The investigation of non-uniform, gradient-dependent releases of molecules from individual cells within the spheroid is left for future work.

\subsection{Received Diffusion Signal by Spheroidal Receiver}
In this subsection, we formulate the Green's functions for boundary value diffusion problems inside and outside the spheroidal receiver in the absence of the spheroidal transmitter. We assume that an impulsive point source transmitter is positioned outside the spheroid at $\bar{r}_0=(r_0,\theta_0,\varphi_0)$, represented by $S(\bar{r}, t, \bar{r}_0, t_0)$. The free-space diffusion \textit{outside the receiving spheroid} is described by the partial differential equation (PDE) \cite{grindrod1996theory}
\begin{align}\label{out_fick_rx}
D{\nabla ^2}c_{\rm orx}(\bar r,t|{\bar{r}_0},{t_0})
+ S(\bar r,t,{\bar{r}_0},t_0) = \frac{{\partial c_{\rm orx}(\bar r,t|{\bar{r}_0},{t_0})}}{{\partial t}}.
\end{align}

In the spherical coordinate system, the Fourier transform of \eqref{out_fick_rx} is re-written as
\small
\begin{align}\label{out_rx_fourier}
&\frac{D}{{{r^2}}}\frac{\partial }{{\partial r}}\left({r^2}\frac{{\partial C_{\rm orx}}}{{\partial r}}\right) +
\frac{D}{{{r^2}\sin \theta }}\frac{\partial }{{\partial \theta }}\left(\sin \theta \frac{{\partial C_{\rm orx}}}{{\partial \theta }}\right)+ \frac{D}{{{r^2}{{\sin }^2}\theta }}\frac{{{\partial ^2}C_{\rm orx}}}{{\partial {\varphi ^2}}}\\\nonumber
&+\frac{{\delta (r  - {r_0})\delta(\theta-\theta_0)\delta(\varphi-\varphi_0)\mathrm{e}^{-i\omega t_0}}}{ r^2\sin\theta  } = i\omega C_{\rm orx}. \nonumber
\end{align}
\normalsize

According to the reaction in \eqref{RC1}, the effective diffusion \textit{inside the receiving spheroid} is modeled as 
\begin{align}\label{fick2rx}
&D_{\rm eff}^{\rm rx}{\nabla ^2}C_{\rm rx}(\bar r,\omega|{\bar{r}_0},{t_0})
- \mathcal K C_{\rm rx}(\bar r,\omega|{\bar{r}_0},{t_0})\\
&   = i\omega C_{\rm rx}(\bar r,\omega|{\bar{r}_0},{t_0}),\nonumber
\end{align}
where $\mathcal K$ is the net \textit{A} molecule reaction rate due to the binding process characterized by the reaction \eqref{RC1}. 
The boundary conditions at the border of the spheroid are given by \eqref{BC2} and \eqref{BC1}.
We have solved the problem and obtained the series-form expression \eqref{Eqsd11} in Appendix \ref{APXA}.
\section{S2S Communication}\label{S2S}
In this section, we first propose the end-to-end diffusive MC channel by determining the molecule concentration received by the spheroidal receiver located at the origin, given the molecules released from the spheroidal transmitter centered at $\bar{r}_{\rm tx}=(r_{\rm tx},\theta_{\rm tx},\varphi_{\rm tx})$. Then, we characterize information transmission between the spheroidal transmitter and the spheroidal receiver in an unbounded environment for an on-off keying modulation scheme to send a sequence of binary data. 
\subsection{S2S Channel Characterization}
We assume that all cells within the spheroidal transmitter release molecules instantaneously at time 0 (i.e., at the beginning of the time slot). This instantaneous release leads to the gradual release of molecules at the boundary of the spheroid, which we approximated with the release function $g(t)$ in \eqref{rlsrate} from the center of the spheroid. Also, we consider ideal aggregate detection across the entire spheroid, where the observations of individual cells are aggregated and used for spheroid signal processing. Finally, we assume that the mutual impact of the spheroidal transmitter and receiver volumes on the individual GFCs is negligible.

Given $c_{\rm rx}(\bar r,t|{{\bar r}_{\rm tx}},{t_0})$ in \eqref{Eqsd11} as the concentration of molecules at the receiver in response to the impulsive point source located at ${{\bar r}_{\rm tx}}$, the concentration profile given the release function $g(t)$ in the spheroidal receiver, $c_{\rm rx}^{\rm Total}(\bar r,t)$, is obtained as
\begin{equation}\label{Ctotal}
c_{\rm rx}^{\rm Total}(\bar r,t)=g(t)\ast c_{\rm rx}(\bar r,t|{\bar r}_{\rm tx},0),
\end{equation} 
where $\ast $ denotes the convolution operator.

Using the GFC in \eqref{Ctotal}, the probability of observing a specific molecule inside a spheroidal receiver at time $t$ is 
\begin{equation}
p_{\rm obs}(t)= \iiint_{\Omega_{\rm RX}}c_{\rm rx}(\bar r,t|{\bar r}_{\rm tx},0)r^2 \sin \theta dr d\theta d\varphi,
\end{equation} 
where $\Omega_{\rm RX}$ represents the set comprising all points within the receiving spheroid. 
The release rate from the transmitting spheroid over $\Delta t$ is modeled as a Poisson random variable (RV) with mean $(g(t) \Delta t)$.

Following \cite{arjmandi2016ion}, we define $\textbf{Y}(t)$ as the number of molecules that are released from the transmitter with release function $g(t)$ within the interval $t \in [0, T_s]$, where $T_s$ is the time slot duration, and subsequently observed at the receiver at time $t$ within the same interval. $\textbf{Y}(t)$  can be modeled using a Poisson distribution with mean
\begin{equation}\label{NOBS}
y(t)=\int_0^t g(t^\prime)p_{\rm obs}(t-t^{\prime})d t^{\prime}=g(t)\ast p_{\rm obs}(t).
\end{equation}
\begin{table}[!t]
\renewcommand{\arraystretch}{1.4}
\caption{Simulation Parameters}
\label{tab:ieee_table}
\centering
\small 
\begin{tabular}{lcc}
\rowcolor{lightgray}\hline
\textbf{Parameter} & \textbf{Variable} & \textbf{Value} \\
\hline
Radius of spheroid & $\gamma_\mathrm{s}$ & $275\,\si{\mu m}$ \\
Number of spheroid's cells  & $N_\mathrm{c}$ & $[1200:4000:24000]$ \\
Volume of one HepaRG cell & $V_\mathrm{c}$ & $3.14\,\mathrm{e}^{-15}\, \si{m^3}$ \\
Reaction rate  & $k_\mathrm{f}$ & 0 or 0.01$\,\si{s^{-1}}$\\
Number of released  molecules  & $N$ & 3000 \\
Time slot & $T_s$& $600$\,\si{s} \\
Diffusion coefficient & $D$ & $1 \times \mathrm{e}^{-9}\, \si{m^2s^{-1}}$ \\
Center of receiver & $\bar{r}_{\rm rx}$ & $(0\, \si{\mu m}, \pi/2, 0)$ \\
Center of transmitter & $\bar{r}_{\rm tx}$ & $(1000\, \si{\mu m}, \pi/2, 0)$ \\
Time step & $\Delta t$& $0.5$\,\si{s} \\
\hline
\end{tabular}
\end{table}
\subsection{Received Signal in On-Off keying modulation}
We adopt the on-off keying modulation scheme where time is divided into time slots of duration $T_s$. In each time slot, the transmission of equiprobable bits 0 and 1 is represented by the average release of 0 and $N$ molecules from \textit{each} cell at the beginning of the time slot, respectively.

At sampling time $t_s$ in each time slot, the received signal is defined as the number of molecules of type $A$ inside the spheroidal receiver volume. The sampling time is chosen to maximize the observation probability $p_{\rm obs}(t)$. The receiver utilizes the observed sample to make a decision regarding the transmitted bit.

We define the sequence of transmitted bits for current and $W$ previous time slots, $B_w = b_w \in [0,1]$, where $w \in [0:W]$. When $w=0$, it corresponds to the current time slot $[0, T_s]$. Conversely, when $w>0$, it indicates the previous time slot $[-wT_s,-(w-1)T_s]$. We immediately have that the number of molecules released at the current time slot ($w=0$) and observed by the receiver at time $t \in [0, T_s]$ is a Poisson-distributed RV with mean $b_0y(t)$ where $y(t)$ is given by \eqref{NOBS}. The number of residual molecules released in the previous time slot, $w$, and observed in the current time slot, $w=0$, at time $t$ is denoted as Poisson RV $\textbf{I}_w(t)$ with mean   
\begin{equation}
I_w(t) =b_wg(wT_s+t)\ast p_{\rm obs}(t).
\end{equation}

The number of molecules observed at time $t \in [0, T_s]$, originating from $W$ previously-transmitted symbols, is given by Poisson RV $\textbf{I}(t)$ with mean 
\begin{equation}
I(t)= \sum_{w=1}^{W}I_w(t)=\sum_{w=1}^{W} b_wg(wT_s+t)\ast p_{\rm obs}(t).
\end{equation}

Therefore, during the current time slot, the receiver's observation at the specific sampling time $t_s$ is given by $\textbf{Y}_R = \textbf{Y}(t_s) + \textbf{I}(t_s)$. In this equation, $\textbf{Y}_R$ is a RV that follows a Poisson distribution with mean $b_0y(t_s)+I(t_s)$.

The performance of our proposed communication system is characterized by its bit error rate. The bit error rate in \eqref{pois2} is derived in Appendix \ref{APXB}.

\section{SIMULATION AND NUMERICAL RESULTS}
\label{results}
In this section, using the parameters provided in Table \ref{tab:ieee_table}, we first evaluate the boundary concentration ratio, $\kappa$, with PBS. We verify the proposed analysis of the transmitting and receiving spheroid Green's functions with this PBS. Additionally, we investigate how the porosity of the transmitting and receiving spheroids affects the molecule concentration inside the receiving spheroid. We compare the received signal at the spheroidal receiver with that of a transparent receiver, revealing the signal amplification and dispersion phenomena within the spheroid. Furthermore, we compare the received signal released from the spheroidal transmitter and the signal released from a point source transmitter.
Finally, the performance of the S2S communication system is evaluated using the BER.
\subsection{System Parameters and Simulation Methods}
The geometric parameters considered in Table 1 are based on the real HepaRG spheroids used in \cite{bauer2017functional} where the spheroid's radius is $275\, \si{\mu m}$ with $N_\mathrm{c}\approx 24000$ cells. The volume of one HepaRG cell is estimated to be $3.14 \times 10^{-15}$\,\si{m^3}. 
The molecules reaction rate is set to be $k_{\mathrm f}=0.01$\,\si{s^{-1}}. Therefore, the $E$ molecule generation rate inside the spheroid is obtained as $\frac{{\partial c_E}}{{\partial t}}=k_\mathrm{f} c_\mathrm{s}$ and correspondingly the term $\mathcal K(\omega)$ in \eqref{fick2rx} is simply $k_\mathrm{f}$.  
\begin{figure}[!t]
	\centering
\includegraphics[width=3.8in]{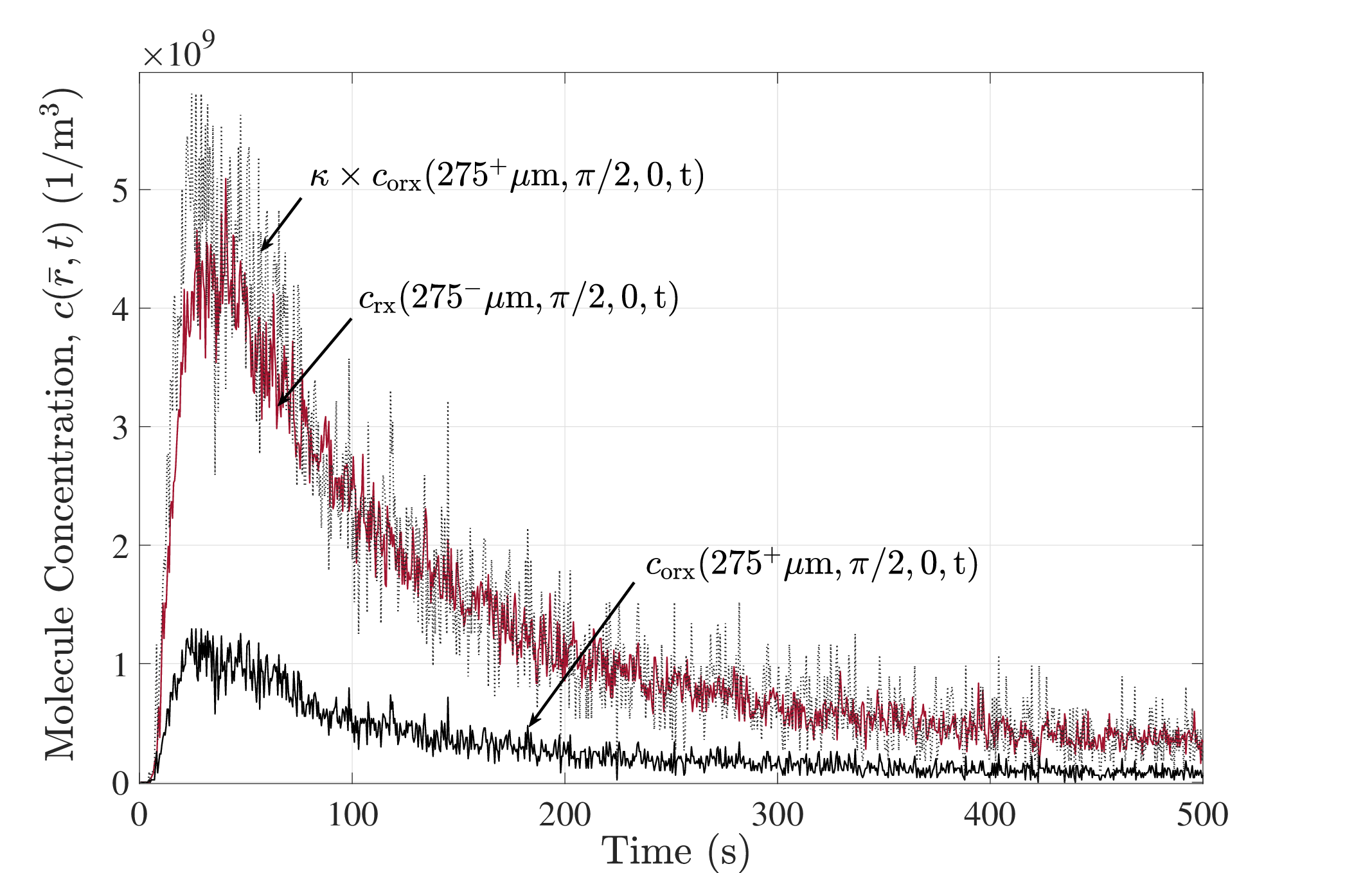}
	\caption{Molecule concentration obtained by PBS at the inside ($c_{\rm rx}$) and outside ($c_{\rm orx}$) of the receiving spheroid boundary ($\bar{r}=(275\,\si{\mu m},\pi/2,0)$) versus time for $N_\mathrm{c}^{\rm rx}=24000$ and $\bar{r}_{\rm tx}=(500\,\si{\mu m},\pi/2,0)$, where $\kappa=\sqrt{\frac{D}{D_{\rm eff}}}$.}
	\label{Figboundary}
\end{figure}

The PBS is implemented in MATLAB (R2021b; The MathWorks, Natick, MA) where the time is divided into time steps of $\Delta t=0.5\,\si{s}$, which is sufficiently small for the results to be insensitive to the time step value. The molecules released from the transmitter move independently in the 3-dimensional space. The displacement of a molecule over $\Delta t$ is modeled as a Gaussian random variable with zero mean and variance $2D\Delta t$ in each Cartesian dimension. For the PBS, the spheroid is simply a diffusion environment with the effective coefficient $D_{\rm eff}$.
When the displacement vector calculated for a molecule outside of the spheroid passes the boundary of the spheroid in a given time slot, the length of the portion of the vector inside the spheroid is updated based on the effective diffusion coefficient $D_{\rm eff}$, i.e., that part of the vector is scaled by the factor $\sqrt{\frac{D_{\rm eff}}{D}}$. Conversely, if the calculated displacement vector of a molecule inside the spheroid in a time slot passes the spheroid boundary, then the length of the portion of the vector outside the spheroid is updated based on the free diffusion coefficient $D$, i.e., that part of the vector is scaled by the factor $\sqrt{\frac{D}{D_{\rm eff}}}$ \cite{arjmandi20233d}. Due to the short time interval ($\Delta t$) and the considerable size of the spheroid, the likelihood of a molecule outside the spheroid entering and leaving the spheroid within a single time interval is low. Additionally, considering the substantial distance between the transmitter and receiver, the direct movement of molecules between spheroids within one time interval is improbable. Nevertheless, we still account for these instances of molecules crossing multiple boundaries within one time interval, where we adjust a molecule's displacement vector at every crossing. Considering the reaction \eqref{RC1} inside the receiving spheroid, a molecule may be absorbed by a cell from the extracellular space of the spheroid during a time step $\Delta t\,\si{s}$ with approximate probability $k_\mathrm{f}\Delta t$ \cite{deng2015modeling}. To represent the homogeneous environment of the spheroidal transmitter, we randomly place $N_\mathrm{c}^{\rm tx}$ cells within the spheroid according to a uniform distribution. To simulate the release from the transmitting spheroid, we assume that each cell within the spheroid releases $N$ molecules simultaneously in response to a stimulus. 
To measure a point concentration at a given time and location, we place a transparent sphere centered at that location with a small radius of $10\,\si{\mu m}$ and count the number of molecules inside the sphere at that time. The concentration would then be the counted number of molecules divided by the volume of the sphere. The concentration value is normalized, i.e., is divided by $N\times N_\mathrm{c}^{\rm tx}$, to obtain the response from the spheroidal transmitter.

\subsection{Single Spheroid Results}
Fig. \ref{Figboundary} shows the concentration at the inside and outside of the outer spheroid boundary, i.e., $c_{\rm rx}(275^- \si{\mu m}, \pi/2, 0,t)$ and $c_{\rm orx}(275^+ \si{\mu m}, \pi/2, 0, t)$, respectively, obtained via PBS given the impulsive source at the transmitter and the porosity determined by $N_\mathrm{c}^{\rm rx}=24000$ cells within the spheroid. We have also plotted the concentration at the outside of the boundary scaled by the factor $\kappa$ to verify the boundary condition \eqref{BC1}. 
The minor mismatch at the peaks is mainly due to the PBS procedure to approximate the concentration at a point. To compute the concentration at the boundary point $(275\,\si{\mu m}, \pi/2, 0)$, we have assumed a transparent sphere of radius $10\; \si{\mu m}$ centered at the boundary point. By partitioning the sphere into left and right hemispheres, we have counted the molecules within each hemisphere. Subsequently, we have normalized the counts by dividing them by the hemisphere volume to approximate the concentration of molecules at the inside and outside of the boundary. The curvature of the boundary causes slight variations in hemisphere volumes, which leads to slightly higher and lower concentration approximations outside and inside the boundary, respectively.

\begin{figure}[!t]
	\centering
	\includegraphics[width=3.8in]{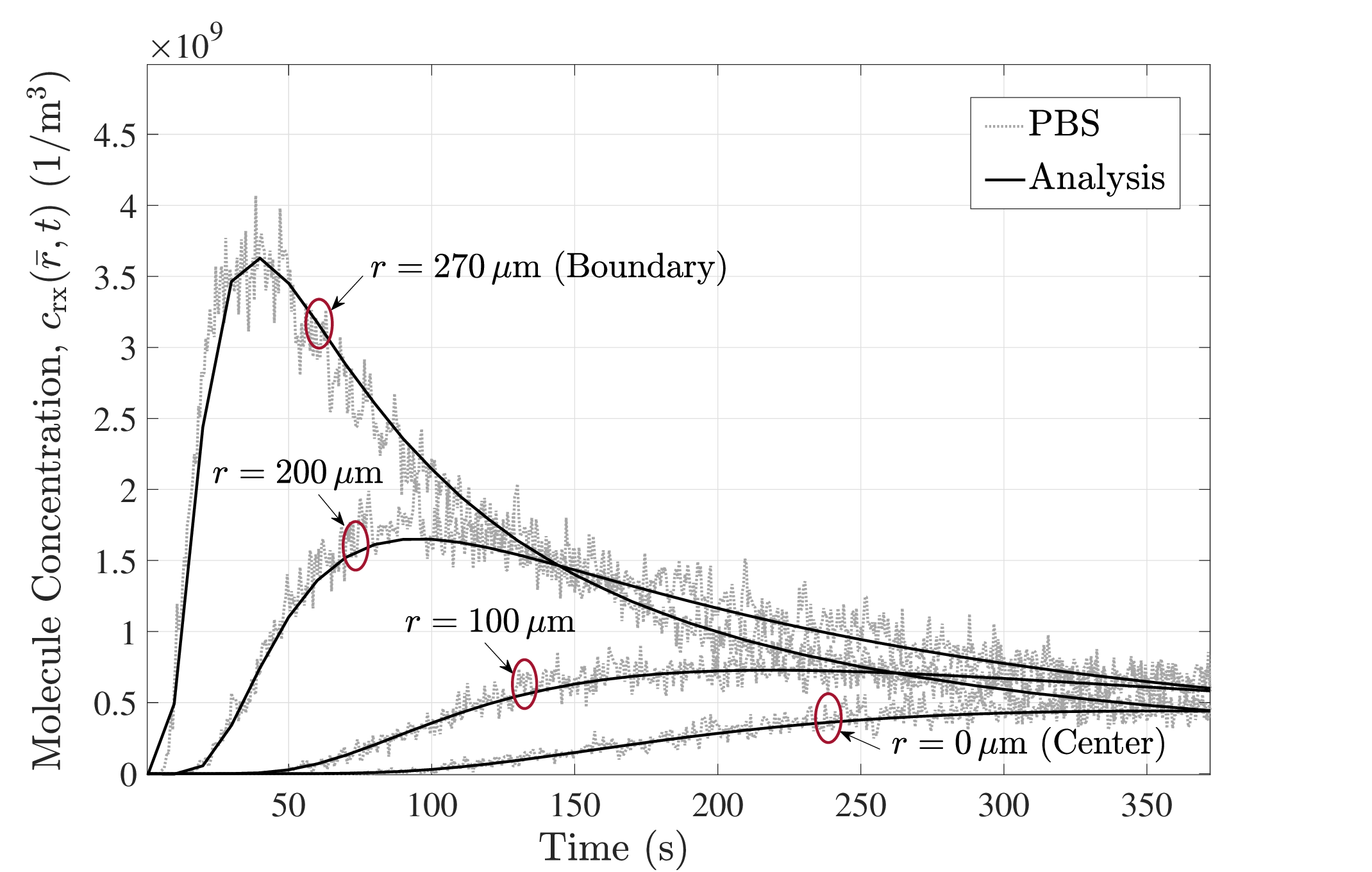}
	\caption{Molecule concentration obtained by analysis and PBS within the spheroidal receiver at different points versus time obtained from analysis and PBS, where $N_\mathrm{c}^{\rm rx}=24000$, $\bar{r}_{0}=(500\,\si{\mu m},\pi/2,0)$, and $\bar{r}= (r,\pi/2,0)$.}
	\label{FigCGF}
\end{figure}

Fig. \ref{FigCGF} depicts the concentration at different points inside the spheroidal receiver given the instantaneous point transmitter at $\bar{r}_0=(500\, \si{\mu m},\pi/2,0)$ obtained from the analysis provided in Appendix \ref{APXA} in \eqref{Eqsd11} and also the PBS when $N_\mathrm{c}^{\rm rx}=24000$. As observed, the PBS fully confirms the analytical results. The figure also indicates that the concentration signal weakens significantly at the points closer to the spheroid center and farther from the point source. In this case, we observe that the peak concentration at the center is about 8 times smaller than that at the boundary. 

According to the findings in Fig. \ref{Figboundary} and Fig. \ref{FigCGF}, as molecules penetrate the porous structure of the spheroid, they become trapped in the pores near the border, which restricts (but does not prevent) their movement inward or outward. This intrinsic spheroid property proves advantageous for MC. The receiver's spheroidal structure enables it to absorb molecules from the environment and keep them within its porous structure for an extended period, which leads to an amplified diffusive signal.

\subsection{End-to-End S2S Results}
\begin{figure}[!t]
	\centering
	\includegraphics[width=3.8in]{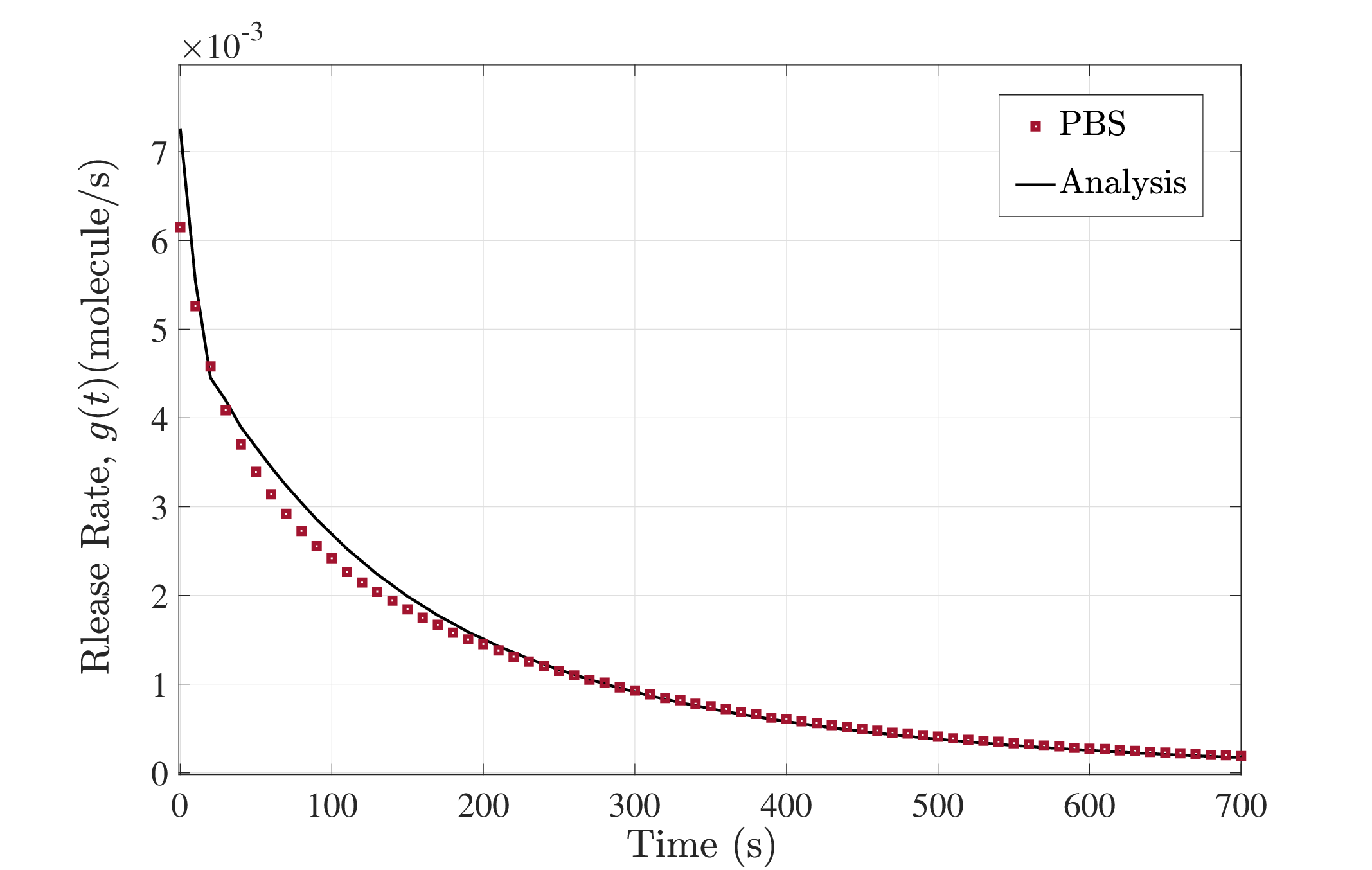}
	\caption{Molecule release rate from the transmitting spheroid for $N_\mathrm{c}^{\rm tx}=24000$.}
	\label{rls rate}
\end{figure}

Fig. \ref{rls rate} shows the molecule release rate from the transmitting spheroid with $N_\mathrm{c}^{\rm tx}=24000$ cells as obtained from \eqref{rlsrate}. As time increases, the release rate decreases since the dynamics of molecules diffusing out of or back into the transmitter are more extreme at the beginning. The analytical release rate was also consistent with the PBS.

\begin{figure}[!t]
	\centering
	\includegraphics[width=3.8in]{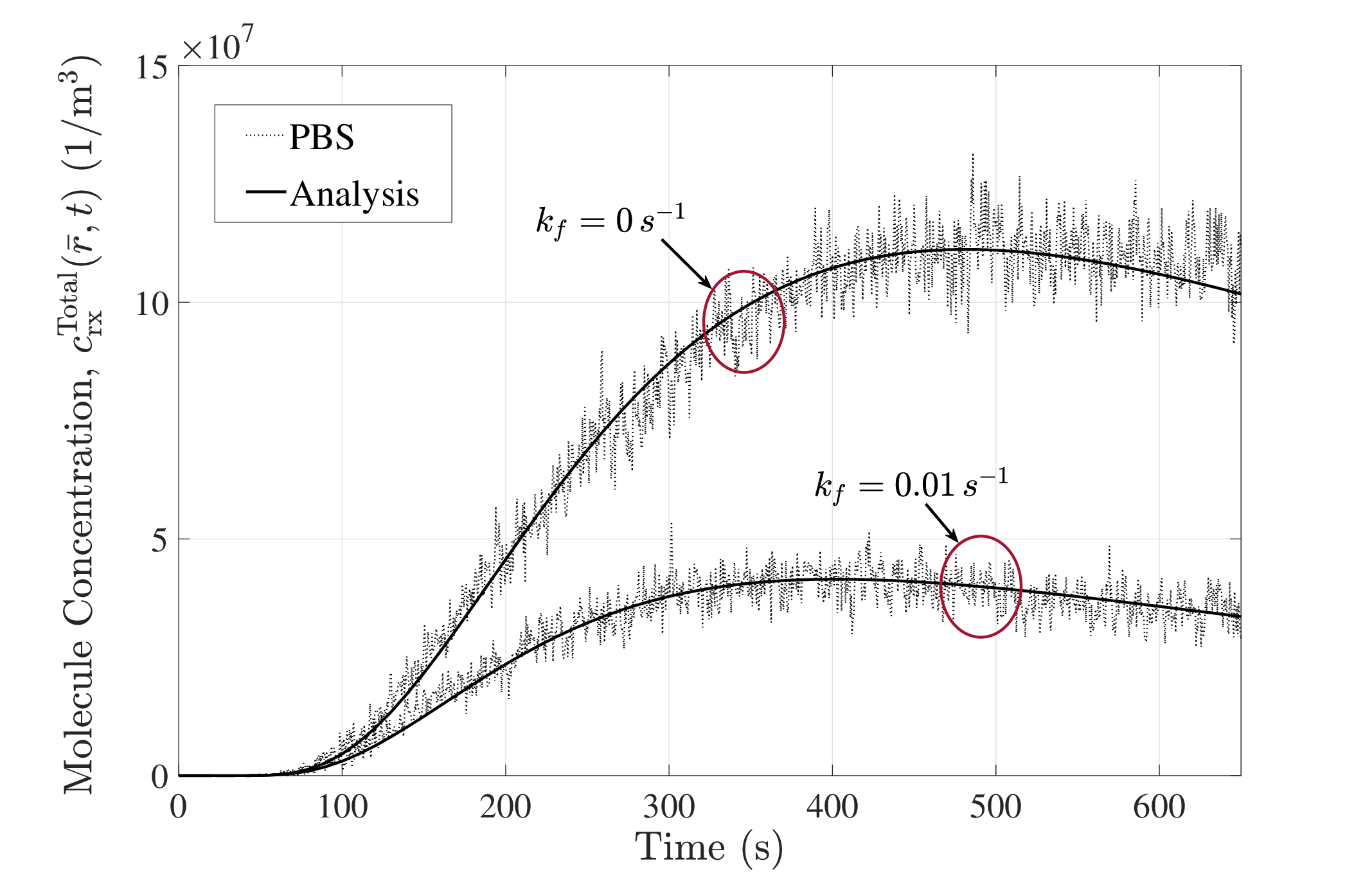}
	\caption{The molecule concentration released from a transmitting spheroid with porosity $\varepsilon_{\rm tx}=0.1349$ at a distance of $1000\, \si{\mu m}$, evaluated at the center of the receiving spheroid with porosity $\varepsilon_{\rm rx}=0.2791$, is compared for two different degradation rates, $k_\mathrm{f}$, over time. The data is obtained through analysis and PBS simulations for $N_\mathrm{c}^{\rm tx}=24000$.}
	\label{centrekd}
\end{figure}

In Fig. \ref{centrekd}, we illustrate the concentration profiles of molecules released from the spheroidal transmitter with release function $g(t)$ and observed in the center of the receiving spheroid with and without degradation as obtained from \eqref{Ctotal}. As observed, our analytical results for the communication between two spheroids are verified with PBS. 
By increasing the degradation rate inside the receiving spheroid, we have a lower concentration. For clarity of exposition, we only consider analytical results for the remaining figures in this paper. 

\begin{figure*}[t]
    \centering
    \begin{minipage}[t]{.48\textwidth}
    \includegraphics[width=\linewidth]{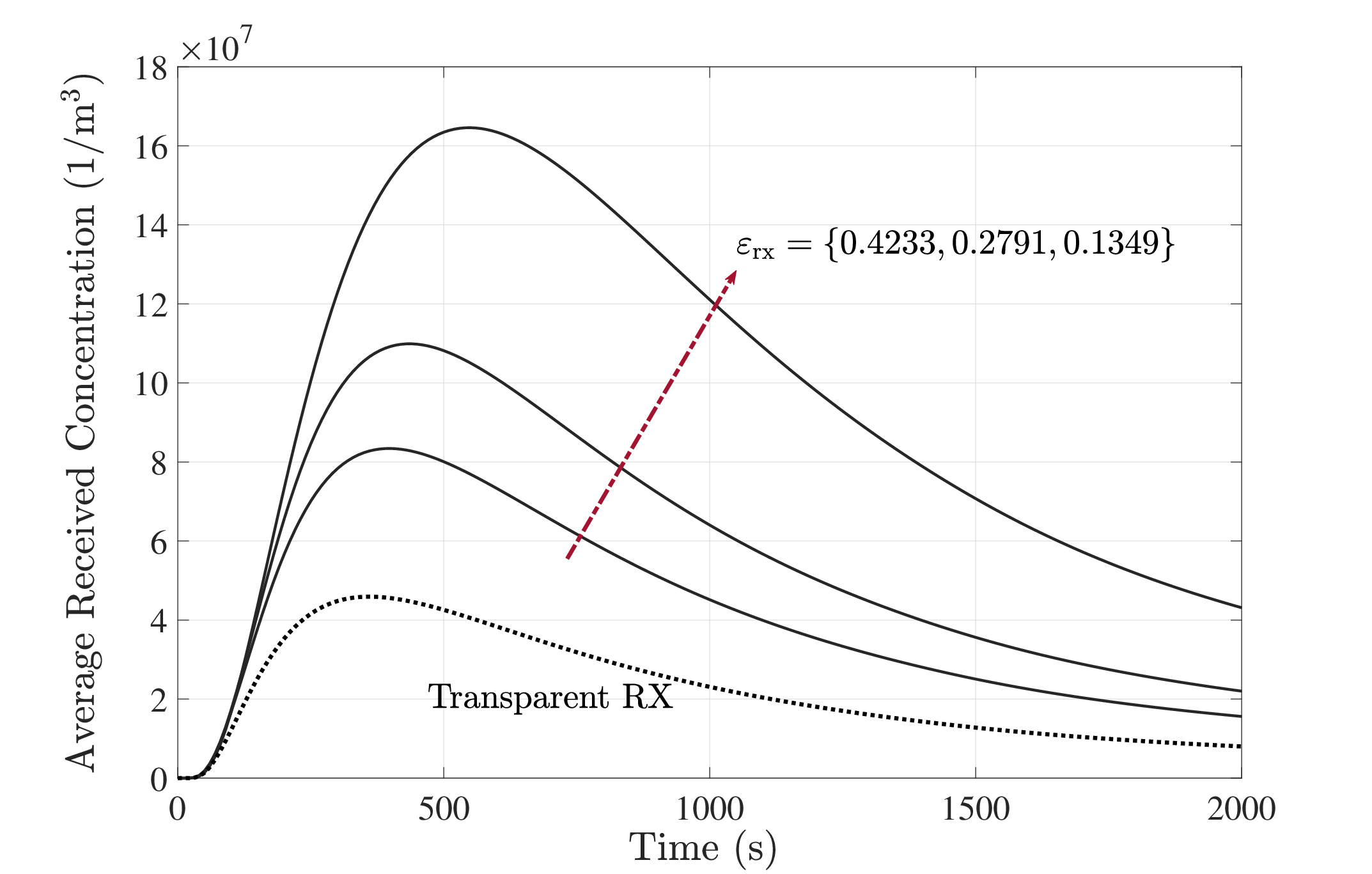}
        \caption{Impact of receiving spheroid porosity on average received concentration over the entire spheroid from a transmitting distance of $1000\,\si{\mu m}$, where $\varepsilon_{\rm tx}=0.1349$. The degradation rates are $k_\mathrm{f}=0$ for the transmitting and receiving spheroids. We also compare the results with the transparent receiver.}
        \label{Fig_Rporosity}
    \end{minipage}\hfill
    \begin{minipage}[t]{.48\textwidth}
    \includegraphics[width=\linewidth]{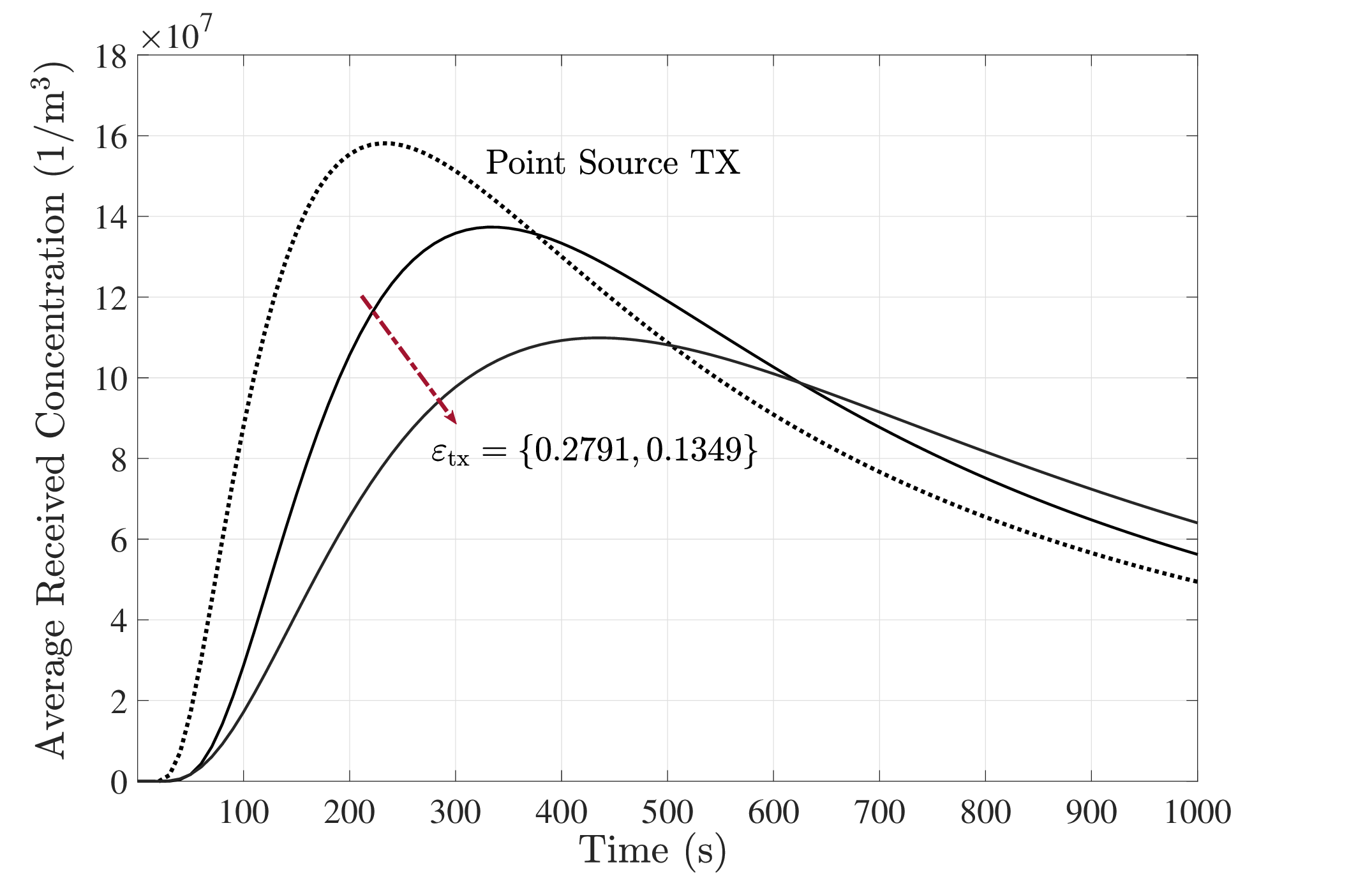}
        \caption{Impact of transmitting spheroid porosity on average received concentration over the entire spheroid from a transmitting distance of $1000\,\si{\mu m}$, where $\varepsilon_{\rm rx}=0.2791$. The degradation rates are $k_\mathrm{f}=0$ for the transmitting and receiving spheroids. We also compare our results with the point source transmitter.}
        \label{3eps.total}
    \end{minipage}
\end{figure*}
We have provided Figs. \ref{Fig_Rporosity} and \ref{3eps.total} to explore the impact of spheroid porosity on signal propagation in the S2S system. In Fig. \ref{Fig_Rporosity}, we keep the porosity level constant in the transmitting spheroid and vary the receiving spheroid porosity via \eqref{eps} by adjusting the number of receiving cells in order to examine how it affects the total concentration of molecules inside the receiver obtained from \eqref{Ctotal}. The results show that reducing the porosity of the receiving spheroid inhibits the rapid molecule diffusion within the porous medium, which leads to a delayed increase in the signal. 
This is because, upon penetration into spheroids with lower porosities, the molecules get stuck inside the more tortuous path between cells and experience a lower effective diffusion coefficient. Their limited ability to escape the porous environment leads to signal amplification. As observed, the spheroidal receiver model displays signal amplification compared to the transparent receiver but at the cost of delay, which can impact the transmission rate.
\begin{figure*}[t]
    \centering
    \begin{minipage}[t]{.48\textwidth}
    \includegraphics[width=\linewidth]{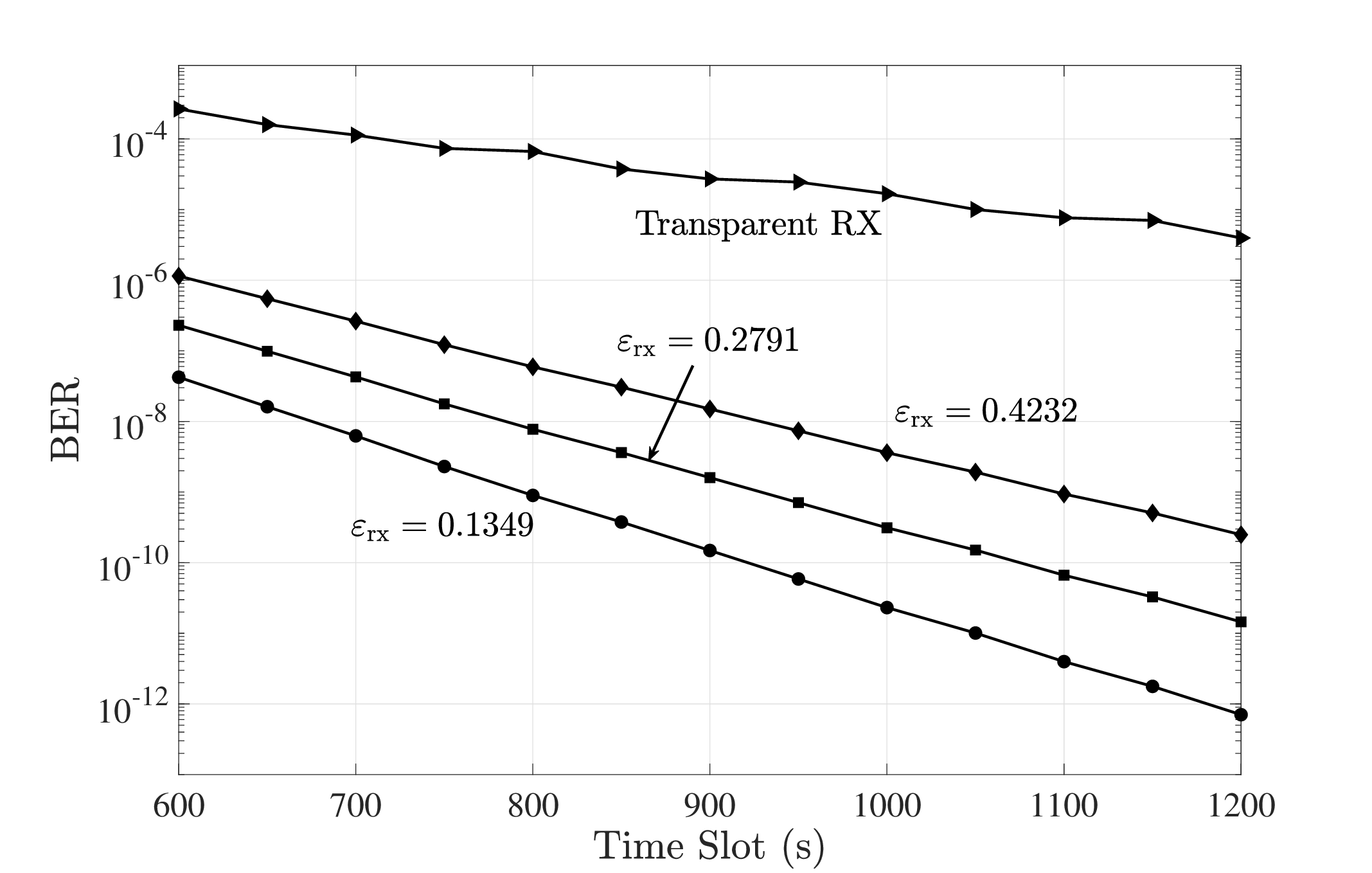}
        \caption{BER of S2S system as a function of $T_s$ for various receiving porosities over a distance of $1000\,\si{\mu m}$, with $\varepsilon_{\rm tx}=0.1349$. The degradation rates are set as $k_\mathrm{f}=0$ for the transmitting spheroid and $k_\mathrm{f}=0.01$ for the receiving spheroid. Each cell within the spheroid releases only one molecule to transmit a bit 1. We also compare the results with the transparent receiver.}
        \label{BER1}
    \end{minipage}\hfill
    \begin{minipage}[t]{.48\textwidth}
    \includegraphics[width=\linewidth]{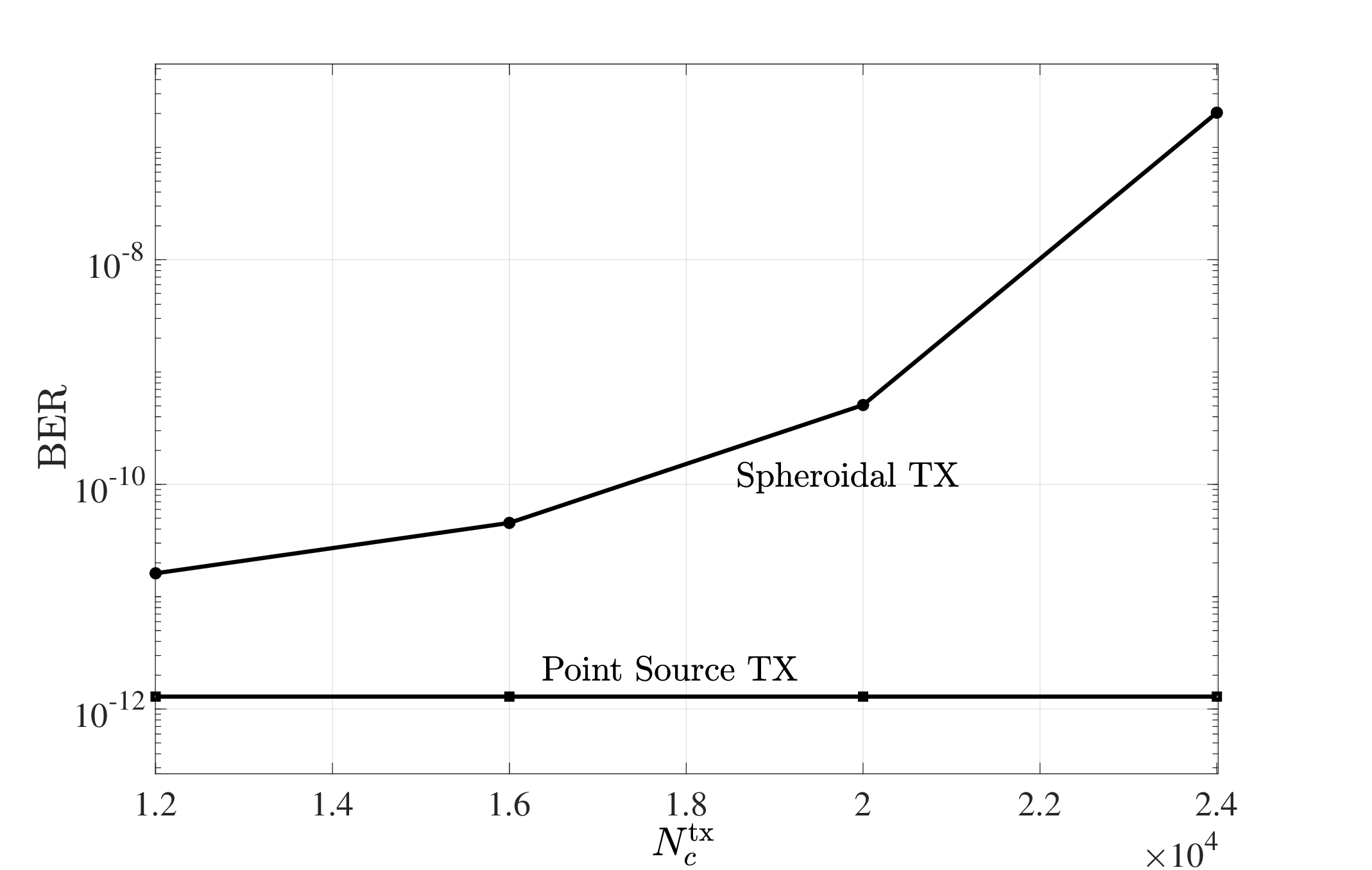}
        \caption{BER for S2S system as a function of the number of cells in the transmitting spheroid over a distance of $1000\,\si{\mu m}$, with $\varepsilon_{\rm rx}=0.2791$. The degradation rates are set as $k_\mathrm{f}=0$ for the transmitting spheroid and $k_\mathrm{f}=0.01$ for the receiving spheroid. The number of released molecules per transmitter is fixed at 24000 molecules. We also compare the results with the point source transmitter.}
        \label{BER2}
    \end{minipage}
\end{figure*}

In Fig. \ref{3eps.total}, we maintain a constant porosity level in the receiving spheroid to examine the influence of the porosity of the transmitting spheroid on the total concentration of molecules inside the receiver as obtained from \eqref{Ctotal}. To isolate the impact of the transmitting spheroid's porosity, we normalize the GFCs by dividing it by the total number of molecules released. Hence, we assume constant ``power'' (i.e., number of molecules released) per transmitter regardless of porosity value. As we increase the porosity level of the transmitter, molecules find it easier to move within the spheroid, causing them to escape more quickly and easily from the porous structure of the spheroid. This leads to a reduction in delay and an increase in the peak of the received signal. Additionally, we compare our spheroidal transmitter with an instantaneous point source transmitter, which indicates the lowest delay and the highest peak signal since the molecules do not get stuck in the porous environment of the transmitting spheroid.

We also evaluate the communications performance of our model by plotting the probability of error in terms of time slot duration for different receiving spheroid porosity in Fig. \ref{BER1}. Given the considerable cell population within the transmitting spheroid, we assume that each cell within the spheroid releases only one molecule to transmit a bit 1 to enable a meaningful performance analysis. 
 Our findings reveal that if we have lower spheroid porosity, as illustrated in Fig. \ref{Fig_Rporosity}, then we have an amplified signal in the receiving spheroid, which can decrease the BER even though the signal is delayed. Furthermore, we conducted a BER comparison between the S2S system and a system where the transmitter is a spheroid and the receiver is transparent. The result demonstrates that incorporating the porous structure for the receiver leads to a lower BER compared to using a transparent receiver. 
The signal amplification caused by the spheroidal structure of the receiver demonstrates effectiveness in improving the accuracy of the detection process within the receiver.

Finally, we measure the impact of transmitting spheroid porosity on the BER in Fig. \ref{BER2} by adjusting the number of transmitting cells. 
By increasing the number of cells or decreasing the porosity, with a fixed number of released molecules per transmitter, we observe a lower molecule concentration in the receiving spheroid with higher delay, as shown in Fig. \ref{3eps.total}. This weakened and delayed signal is associated with a subsequent increase in the BER. The figure also compares the BER between the S2S system and a point source transmitter releasing 24000 molecules per bit 1. This analysis underscores that as the porosity of the transmitting spheroid decreases in the S2S system, the system's performance approaches that of a system comprised of a point source transmitter and a spheroidal receiver.

\section{Conclusion} \label{Conclusion}
We proposed an end-to-end diffusive molecular communication (MC) system using a spheroidal transmitter and receiver.
We successfully derived the analytical solution for the Green's function for concentration (GFC) for both transmitter and receiver spheroids.
As confirmed by PBS, we investigated how different porosity values in both the spheroidal transmitter and receiver impact the GFC and consequently the received signal. Our findings indicate that increasing the porosity of the receiving spheroid leads to a lower peak GFC with a reduced delay.  In addition, we observed that increasing the porosity of the transmitting spheroid while maintaining a constant number of released molecules from the entire spheroid results in a higher concentration peak inside the receiver and a shorter delay. The evaluation of our proposed system's performance is conducted using the bit error rate (BER) in terms of time slot duration and transmitter porosity under the effect of inter-symbol interference (ISI).
The results show that lower porosity in the receiving spheroid, although causing signal delay, leads to reduced BER since it amplifies the concentration, which can enhance the detection accuracy in the receiver. However, by decreasing the porosity of the transmitting spheroid while keeping the transmission power constant, the receiving spheroid experiences a decrease in molecule concentration, resulting in a higher BER.

This paper represents an initial step toward an MC system that employs a transmitter and receiver model based on aggregations of cells. The results of this research will create a solid foundation for future advancements in the field of MC. In future work, one might improve model accuracy or examine signal diffusion in a nonhomogeneous porous media model of the spheroid, for example by including (i) the necrotic core that can result from cell death in the center of a spheroid due to a lack of oxygen; or (ii) a spheroid of two (or more) different cell types, e.g., HepaRG cells and stellate cells. Another challenge is to perform experimental validation, which would require measurements of internal spheroid concentration. Future studies can consider extending the model to more than two spheroids to illustrate more realistic scenarios. Considerations can also include modeling spheroidal transceivers within bounded microfluidic wells, accounting for fluid flow dynamics that could promote deeper molecule penetration into the spheroid.  

\appendices
\section{Transmitter and Receiver Analysis}\label{APXA}
\begin{figure*}
\begin{equation}\label{matchtx}
	\begin{array}{l}
\sum_{n=0}^{\infty}\sum_{m=0}^{n} H_{mn} \frac{D_{\rm eff}^{\rm tx}}{{{r^2}}}\frac{\partial }{{\partial r}}({r^2}\frac{{\partial R_n^{\rm tx}(r,\omega)}}{{\partial r}}) P_n^m(\cos\theta)\cos(m(\varphi-\varphi_{0})) +\sum_{n=0}^{\infty}\sum_{m=0}^{n} H_{mn} R_n^{\rm tx}(r,\omega) \frac{D_{\rm eff}^{\rm tx}}{{{r^2}\sin \theta }}\frac{\partial }{{\partial \theta }}(\sin \theta \frac{{\partial P_n^m(\cos\theta)}}{{\partial \theta }}) +\\ \frac{D_{\rm eff}^{\rm tx}}{{{r^2}{{\sin }^2}\theta }}\frac{{{\partial ^2}(\cos(m(\varphi-\varphi_{0})))}}{{\partial {\varphi ^2}}}
	-({i\omega})\sum_{n=0}^{\infty}\sum_{m=0}^{n} H_{mn}R_n^{\rm tx}(r,\omega) P_n^m(\cos\theta)\cos(m(\varphi-\varphi_{0}))\\
	= 	-\sum\limits_{n=0}^\infty  \sum\limits_{m = 0}^n 
	L_m \frac{{2n + 1}}{2}\frac{{(n - m)!}}{{(n + m)!}}
	\times{P_n^m(\cos \theta )P_n^m(\cos {\theta _{0}})} {\cos (m(\varphi  - \varphi _{0}))}\delta(r-r_0).
	\end{array}
	\end{equation}
	\color{black}
	\hrulefill
\end{figure*}
\begin{figure*}
\begin{equation}\label{match2tx}
\begin{array}{l}
\sum_{n=0}^{\infty}\sum_{m=0}^{n} H_{mn} \frac{D}{{{r^2}}}\frac{\partial }{{\partial r}}({r^2}\frac{{\partial R_n^{\rm otx}(r,\omega)}}{{\partial r}}) P_n^m(\cos\theta)\cos(m(\varphi-\varphi_{0})) +\sum_{n=0}^{\infty}\sum_{m=0}^{n} H_{mn} R_n^{\rm otx}(r,\omega) \frac{D}{{{r^2}\sin \theta }}\frac{\partial }{{\partial \theta }}(\sin \theta \frac{{\partial P_n^m(\cos\theta)}}{{\partial \theta }}) +\\ \frac{D}{{{r^2}{{\sin }^2}\theta }}\frac{{{\partial ^2}(\cos(m(\varphi-\varphi_{0})))}}{{\partial {\varphi ^2}}}
	-({i\omega})\sum_{n=0}^{\infty}\sum_{m=0}^{n} H_{mn}R_n^{\rm otx}(r,\omega) P_n^m(\cos\theta)\cos(m(\varphi-\varphi_{0}))= 0.
	\end{array}
	\end{equation}
	\color{black}
	\hrulefill
\end{figure*}
\begin{figure*}
\begin{equation}\label{Cotx}
	\begin{bmatrix}
	0 & j_n(k^{\rm tx} \gamma_{\rm tx}) & y_n(k^{\rm tx} \gamma_{\rm tx}) & -\kappa_{\rm tx} h_n(k^{\rm otx} \gamma_{\rm tx})\\
	0 & D_{\rm eff}^{\rm tx} k^{\rm tx} j'_n(k^{\rm tx}\gamma_{\rm tx}) & D_{\rm eff}^{\rm tx}k^{\rm tx} y'_n(k^{\rm tx}\gamma_{\rm tx}) & -D k^{\rm otx} h'_n(k^{\rm otx}\gamma_{\rm tx})\\
	j_n(k^{\rm tx}r_0) & -j_n(k^{\rm tx}r_0) & -y_n(k^{\rm tx}r_0) & 0\\
	r^2_{ 0}k^{\rm tx}j'_n(k_1r_0) & -r^2_{0}k^{0} j'_n(k^{\rm tx}r_0) & -r^2_{0}y'_n(k^{\rm tx}r_0) & 0
	\end{bmatrix}
	\begin{bmatrix}
	G_n\\
	A_n\\
	B_n\\
	D_n
	\end{bmatrix}
	=
	\begin{bmatrix}
	0\\
	0\\
	0\\
	1
	\end{bmatrix}.
	\end{equation}
		\color{black}
	\hrulefill
\end{figure*}
\balance
In the following, we obtain analytical expressions of Green's function for both the spheroidal transmitter and receiver in the frequency domain. We consider homogeneous initial conditions for the system.
The Green's function for the boundary value diffusion problem for inside and outside of the spheroidal transmitter is given by \eqref{tx_fick_fourier} and \eqref{out_fick_fourier}, given the boundary conditions \eqref{BC2} and \eqref{BC1} by replacing $c_\mathrm{s}$ and $c_\mathrm{o}$ with $c_{\rm tx}$ and $c_{\rm otx}$, respectively. Also, the Green's function for the boundary value diffusion problem for outside and inside of the spheroidal receiver is given by \eqref{out_fick_rx} and \eqref{fick2rx}, given the boundary conditions \eqref{BC2} and \eqref{BC1} by replacing $c_\mathrm{s}$ and $c_\mathrm{o}$ with $c_{\rm rx}$ and $c_{\rm orx}$, respectively.
\subsection{General Analytical Solution}
The general solution of both problems is considered as \cite{zoofaghari2021modeling} 
\begin{equation}\label{Eqsd11}
\begin{aligned}
C_q(r,\theta ,\varphi ,\omega|\bar{r}_0)
& = \sum\limits_{n = 0}^\infty{\sum\limits_{m = 0}^n {{{H_{mn}R_n^q(r,\omega)}\cos (m(\varphi  - {\varphi _{\rm tx}}))} } } 
\\  
&\times P_n^m(\cos \theta ),\;\; q\in\{\rm tx,otx,orx,rx\},
\end{aligned}
\end{equation} 
where  $R_n^q(r,\omega)$  is the unknown radial Green function of $r$ for the region $q\in\{\rm tx, otx, orx, rx\}$, $\mathcal{F}(\theta,\varphi)=\cos(m \varphi)P_n^m(\cos\theta)$ is a spherical harmonic with degree $n$ and order $m$ that  satisfies the  partial differential equation (PDE) in
\begin{equation}\label{logender22}
\begin{aligned}
&\frac{D}{{\sin \theta }}\frac{\partial }{{\partial \theta }}\left(\sin \theta \frac{{\partial \mathcal{F}(\theta,\varphi)}}{{\partial \theta }}\right) + \frac{D}{{{{\sin }^2}\theta }}\frac{{{\partial ^2}\mathcal{F}(\theta,\varphi)}}{{\partial {\varphi ^2}}}\\
&+{n(n+1)}\mathcal{F}(\theta,\varphi)=0,
\end{aligned}
\end{equation}
and $H_{mn}$ denotes the unknown coefficient that needs to be determined for the particular configuration. 
Also, the representation of
${\delta (\varphi  - {\varphi _0})} \frac{{\delta (\theta  - {\theta _0})}}{{\sin \theta }}$  based on the aforementioned spherical harmonic is given by  
\begin{equation}\label{deltaphitheta1}
\begin{aligned}
&{\delta (\varphi  - {\varphi _0})} \frac{{\delta (\theta  - {\theta _0})}}{{\sin \theta }} =\sum\limits_{n=0}^\infty  \sum\limits_{m = 0}^n L_m \frac{{2n + 1}}{2}\frac{{(n - m)!}}{{(n + m)!}}\\
&\times {P_n^m(\cos \theta )P_n^m(\cos {\theta _0})} {\cos \big(m(\varphi  - \varphi _0)\big)},
\end{aligned}
\end{equation}
where $L_0=\frac{1}{2\pi}$ and $L_m=\frac{1}{\pi}, m\geq 1$.

\subsection{Transmitter Region}
Replacing $C_q$ and ${\delta (\varphi  - {\varphi _0})} \frac{{\delta (\theta  - {\theta _0})}}{{\sin \theta }}$  in \eqref{tx_fick_fourier} and \eqref{out_fick_fourier} by the corresponding series-form representations given by \eqref{Eqsd11} and \eqref{deltaphitheta1}, respectively, leads to \eqref{matchtx} and \eqref{match2tx} at the top of the next page. Matching the two sides of \eqref{matchtx} yields
\begin{equation}
H_{mn}=L_m \frac{{2n + 1}}{2}\frac{{(n - m)!}}{{(n + m)!}}P_n^m(\cos\theta_0),
\end{equation} 
and
\begin{equation}\label{RPDEtx}
\begin{aligned}
&r^2\frac{\partial^2 R_n^{\rm tx}(r,\omega)}{\partial r^2}+2r\frac{\partial R_n^{\rm tx}(r,\omega)}{\partial r}\\
&+((k^{\rm tx})^2r^2-n(n+1))R_n^{\rm tx}(r,\omega)=\delta ( r -r_0),
\end{aligned}
\end{equation}
where $k^{\rm tx}=\sqrt{\frac{{\mathcal -i\omega}}{D^{\rm tx}_{\rm eff}}}$. Similarly, \eqref{match2tx} is reduced to
\begin{equation}\label{RPDE2tx}
\begin{aligned}
&r^2\frac{\partial^2 R_n^{\rm otx}(r,\omega)}{\partial r^2}+2r\frac{\partial R_n^{\rm otx}(r,\omega)}{\partial r}\\
&+((k^{\rm otx})^2r^2-n(n+1))R_n^{\rm otx}(r,\omega)=0,
\end{aligned}
\end{equation}
where $k^{\rm otx}=\sqrt{\frac{-i\omega}{D}}$.

By applying \eqref{Eqsd11} in the Fourier forms of  boundary conditions \eqref{BC2} and \eqref{BC1}, we obtain 
\begin{align}
D^{\rm tx}_{\rm eff}\frac{\partial R_n^{\rm tx}(r,\omega)}{\partial r }\bigg|_{ r=\gamma_{\rm tx}^{-}} &= D\frac{\partial R_n^{\rm otx}(r,\omega)}{\partial r }\bigg|_{r =\gamma_{\rm tx}^{+}} \label{B1tx}, \\
R_n^{\rm tx}(r,\omega)\bigg|_{r =\gamma_{\rm tx}^{-}} &= \kappa_{\rm tx}\times R_n^{\rm otx}(r,\omega)\bigg|_{r =\gamma_{\rm tx}^{+}}, \label{B2tx}
\end{align}
where $\kappa_{\rm tx}=\sqrt{\frac{D}{D_{\rm eff}^{\rm tx}}}$ is the concentration ratio at the boundary of the transmitting spheroid. In these equations, superscripts + and - represent a minuscule distance above or below the transmitter radius $\gamma_{\rm tx}$, respectively.

To establish the discontinuity boundary condition of the Green's function at the point source location $r = r_0$, we integrate \eqref{RPDEtx} across a small interval $[r_{0}^{-}=r_0-\epsilon,r_{0}^{+}=r_0+\epsilon]$. Taking the limit $\epsilon \rightarrow 0$ and employing the sifting property of the Dirac delta function at the right-hand side leads to the boundary condition \cite{zoofaghari2021modeling}
\begin{equation}\label{SDtx}
r^2 \frac{\partial R_n^{\rm tx}(r,\omega)}{\partial r}\bigg|_{r=r_{0}^{+}}
- r^2 \frac{\partial R_n^{\rm tx}(r,\omega)}{\partial r}\bigg|_{r=r_{0}^{-}}=1.
\end{equation}

The solutions of the homogeneous form of the PDE of \eqref{RPDEtx} and also \eqref{RPDE2tx}, intended for the transmitter, are then given by
\begin{align}
R_n^{\rm tx}(r,\omega)& = \left\{ {\begin{array}{*{20}{c}}
	{\begin{array}{*{20}{c}}
		G_nj_n(kr),\;{r<r_0}
		\end{array}}\label{Rrtx}\\
	{\begin{array}{*{20}{c}}
		A_nj_n(kr)+B_n y_n(kr),\;{r_0<r<\gamma_{\rm tx}}
		\end{array}}
	\end{array}} \right.,\\
R_n^{\rm otx}(r,\omega)& = D_nh_n(kr), \; r>\gamma_{\rm tx},\label{Rrtx01}
\end{align}
where $j_n(\cdot)$ and $y_n(\cdot)$ are the $n$th order of the first and second kinds of spherical Bessel function, respectively, and $h_n(\cdot)$ represents the $n$th order Hankel function of the first type. 
By applying the solutions \eqref{Rrtx} and \eqref{Rrtx01} to the boundary conditions \eqref{B1tx}, \eqref{B2tx}, and \eqref{SDtx}, and also  the continuity condition of concentration  at ${\bar{r}_{0}}$, i.e., 
\begin{equation}\label{SC}
R_n^{\rm tx}(r,\omega)\bigg|_{r=r^{+}_{0}}
= R_n^{\rm tx}(r,\omega)\bigg|_{r=r^{-}_{0}},
\end{equation} 
the system of linear equations \eqref{Cotx} is obtained at the top of this page and from which the coefficients $A_n$, $B_n$, $G_n$, and $D_n$ are calculated.
We note that $j'(\cdot)$ and $y'(\cdot)$ in \eqref{Cotx} are the derivatives of the functions $j(\cdot)$ and $y(\cdot)$ with respect to $r$, respectively.  By having the coefficients $A_n,B_n, C_n$, and $D_n$, $R_n^q(r,\omega)$, $q\in\{\rm tx, otx\}$ in \eqref{Rrtx} is known and the GFCs of the inside and outside of the transmitting spheroid are computed by taking the inverse Fourier transform of \eqref{Eqsd11} for the inside and outside, respectively.

\subsection{Receiver Region}
\begin{figure*}
	\begin{equation}\label{matchrx}
	\begin{array}{l}
	\sum_{n=0}^{\infty}\sum_{m=0}^{n} H_{mn} \frac{D}{{{r^2}}}\frac{\partial }{{\partial r}}({r^2}\frac{{\partial R_n^{\rm orx}(r,\omega)}}{{\partial r}}) P_n^m(\cos\theta)\cos(m(\varphi-\varphi_{0})) +\sum_{n=0}^{\infty}\sum_{m=0}^{n} H_{mn} R_n^{\rm orx}(r,\omega) \frac{D}{{{r^2}\sin \theta }}\frac{\partial }{{\partial \theta }}(\sin \theta \frac{{\partial P_n^m(\cos\theta)}}{{\partial \theta }}) +\\ \frac{D}{{{r^2}{{\sin }^2}\theta }}\frac{{{\partial ^2}(\cos(m(\varphi-\varphi_{0})))}}{{\partial {\varphi ^2}}}
	-({i\omega})\sum_{n=0}^{\infty}\sum_{m=0}^{n} H_{mn}R_n^{\rm orx}(r,\omega) P_n^m(\cos\theta)\cos(m(\varphi-\varphi_{0}))\\
	= 	-\sum\limits_{n=0}^\infty  \sum\limits_{m = 0}^n 
	L_m \frac{{2n + 1}}{2}\frac{{(n - m)!}}{{(n + m)!}}
	\times{P_n^m(\cos \theta )P_n^m(\cos {\theta _{0}})} {\cos (m(\varphi  - \varphi _{0}))}\delta(r-r_0).
	\end{array}
	\end{equation}
	\color{black}
	\hrulefill
\end{figure*}
\begin{figure*}
	\begin{equation}\label{match2rx}
	\begin{array}{l}
\sum_{n=0}^{\infty}\sum_{m=0}^{n} H_{mn} \frac{D}{{{r^2}}}\frac{\partial }{{\partial r}}({r^2}\frac{{\partial R_n^{\rm rx}(r,\omega)}}{{\partial r}}) P_n^m(\cos\theta)\cos(m(\varphi-\varphi_{0}) +\sum_{n=0}^{\infty}\sum_{m=0}^{n} H_{mn} R_n^{\rm rx}(r,\omega) \frac{D}{{{r^2}\sin \theta }}\frac{\partial }{{\partial \theta }}(\sin \theta \frac{{\partial P_n^m(\cos\theta)}}{{\partial \theta }}) +\\ \frac{D}{{{r^2}{{\sin }^2}\theta }}\frac{{{\partial ^2}(\cos(m(\varphi-\varphi_{0})))}}{{\partial {\varphi ^2}}}
	-(\mathcal K(i\omega)+{i\omega})\sum_{n=0}^{\infty}\sum_{m=0}^{n} H_{mn}R_n^{\rm rx}(r,\omega) P_n^m(\cos\theta)\cos(m(\varphi-\varphi_{0}))= 	0.
	\end{array}
	\end{equation}
	\color{black}
	\hrulefill
\end{figure*}
\begin{figure*}
	\begin{equation}\label{Corx}
	\begin{bmatrix}
	j_n(k^{\rm rx}\gamma_{\rm rx}) & -\kappa_{\rm rx} y_n(k^{\rm orx}\gamma_{\rm rx}) & -\kappa_{\rm rx}j'_n(k^{\rm orx}\gamma_{\rm rx}) & 0\\
	D_{\rm eff} k^{\rm rx} j_n(k^{\rm rx}\gamma_{\rm rx}) & -Dk^{\rm orx}y'_n(k^{\rm orx}\gamma_{\rm rx}) & -Dk^{\rm orx}j'_n(k^{\rm orx}\gamma_{\rm rx}) & 0\\
	0 & y_n(k^{\rm orx}r_0) & j_n(k^{\rm orx}r_0) & -h_n(k^{\rm orx}r_0)\\
	0 & r^2_0y'_n(k^{\rm orx}r_0) & r^2_0 j'_n(k^{\rm orx}r_0) & -r^2_0h'_n(k^{\rm orx}r_0)
	\end{bmatrix}
	\begin{bmatrix}
	G_n\\
	A_n\\
	B_n\\
	D_n
	\end{bmatrix}
	=
	\begin{bmatrix}
	0\\
	0\\
	0\\
	1
	\end{bmatrix}.
	\end{equation}
\end{figure*}
Similar to the transmitter case, replacing $C_q$ and ${\delta (\varphi  - {\varphi _0})} \frac{{\delta (\theta  - {\theta _0})}}{{\sin \theta }}$  in \eqref{out_rx_fourier} and \eqref{fick2rx} by the corresponding series-form representations given by \eqref{Eqsd11} and \eqref{deltaphitheta1}, respectively, leads to \eqref{matchrx} and \eqref{match2rx} at the top of the next page. Matching the two sides of \eqref{matchrx} yields
\begin{equation}
H_{mn}=L_m \frac{{2n + 1}}{2}\frac{{(n - m)!}}{{(n + m)!}}P_n^m(\cos\theta_0),
\end{equation} 
and
\begin{equation}\label{RPDE}
\begin{aligned}
&r^2\frac{\partial^2 R_n^{\rm orx}(r,\omega)}{\partial r^2}+2r\frac{\partial R_n^{\rm orx}(r,\omega)}{\partial r}\\
&+((k^{\rm rx})^2r^2-n(n+1))R_n^{\rm orx}(r,\omega)=\delta (r - r_{0}),
\end{aligned}
\end{equation}
where  $k^{\rm rx}=\sqrt{\frac{-i\omega}{D}}$.
Similarly, \eqref{match2rx} is reduced to
\begin{equation}\label{RPDE2}
\begin{aligned}
&r^2\frac{\partial^2 R_n^{\rm rx}(r,\omega)}{\partial r^2}+2r\frac{\partial R_n^{\rm rx}(r,\omega)}{\partial r}\\
&+((k^{\rm orx})^2r^2-n(n+1))R_n^{\rm rx}(r,\omega)=0,
\end{aligned}
\end{equation}
where  $k^{\rm orx}=\sqrt{\frac{-{\mathcal K(i\omega)-i\omega}}{D_{\rm eff}^{\rm rx}}}$. We emphasize that $k^{\rm orx}$ is a function of the degradation rate via $\mathcal K$.

Similar to the transmitter scenario, when we apply \eqref{Eqsd11} in the Fourier forms of  boundary conditions \eqref{BC2} and \eqref{BC1}, above or below the receiver radius $\gamma_{\rm rx}$, we obtain 
\begin{align}
D_{\rm eff}^{\rm rx}\frac{\partial R_n^{\rm rx}(r,\omega)}{\partial r }\bigg|_{ r=\gamma_{\rm rx}^-} &= D\frac{\partial R_n^{\rm orx}(r,\omega)}{\partial r }\bigg|_{r =\gamma_{\rm rx}^+} \label{B1}, \\
R_n^{\rm rx}(r,\omega)\bigg|_{r =\gamma_{\rm rx}^-} &= \kappa_{\rm rx}\times R_n^{\rm orx}(r,\omega)\bigg|_{r =\gamma_{\rm rx}^+}, \label{B2}
\end{align}
where $\kappa_{\rm rx}=\sqrt{\frac{D}{D_{\rm eff}^{\rm rx}}}$ is the concentration ratio at the boundary of the receiving spheroid. 


Similar to what we did to obtain \eqref{SDtx}, we integrate \eqref{RPDE} across a small interval below and above the point source location $r=r_0$, which leads to the corresponding boundary condition \cite{zoofaghari2021modeling}
\begin{equation}\label{SD}
r^2 \frac{\partial R_n^{\rm orx}(r,\omega)}{\partial r}\bigg|_{r=r^{+}_{0}}
- r^2 \frac{\partial R_n^{\rm orx}(r,\omega)}{\partial r}\bigg|_{r=r^{-}_{0}}=1.
\end{equation}

The solutions of the homogeneous form of the PDE of \eqref{RPDE} and also \eqref{RPDE2}, applicable to the receiver region, are then given by
\begin{align}
R_n^{\rm rx}(r,\omega) & = G_nj_n(kr), \; r<\gamma_{\rm rx},\label{Rr} \\ 
R_n^{\rm orx}(r,\omega)& = \left\{ {\begin{array}{*{20}{c}}
	{\begin{array}{*{20}{c}}
		B_nj_n(kr)+A_n y_n(kr),\;{\gamma_{\rm rx}<r<r_0}
		\end{array}}\\ \label{Rr01}
	{\begin{array}{*{20}{c}}
		D_nh_n(kr),\;{r>r_0}
		\end{array}}
	\end{array}}\right.  .
\end{align}

By applying the solutions \eqref{Rr} and \eqref{Rr01} to the boundary conditions \eqref{B1}, \eqref{B2}, and \eqref{SD}, and also  the continuity condition of concentration  at $\bar{r}_{0}$, i.e., 
\begin{equation}\label{SC}
R_n^{\rm orx}(r,\omega)\bigg|_{r=r^{+}_{0}}
= R_n^{\rm orx}(r,\omega)\bigg|_{r=r^{-}_{0}},
\end{equation} 
the system of linear equations \eqref{Corx} is obtained at the top of this page and by which the coefficients $A_n$, $B_n$, $G_n$, and $D_n$ are calculated.
By having the coefficients $A_n,B_n, G_n$, and $D_n$, $R_n^q(r,\omega)$, $q\in\{\rm rx, orx\}$ in \eqref{Rr} is known and the GFCs of the inside and outside of the receiving spheroid are computed by taking the inverse Fourier transform of \eqref{Eqsd11} for the inside and outside, respectively.
\section{On-Off Keying S2S Performance Analysis} \label{APXB}
The RV $\textbf{Y}_R$, which follows a Poisson distribution, can be described in the following manner:      
\begin{align}\label{pois1}
&  \Pr(\textbf{Y}_R=y|b_0,b_1,...,b_W)\nonumber\\
&   =\dfrac{\mathrm{e}^{-\mathbb{E}((\textbf{Y}_R|b_0,b_1,...,b_W))}\left(-\mathbb{E}((\textbf{Y}_R|b_0,b_1,...,b_W)))\right)^y}{y!},
\end{align}
in which
\begin{align}\label{pois2}
\mathbb{E}((\textbf{Y}_R|b_0,b_1,...,b_W))& =\sum_{i=0}^{W}b_ig(iT_s+t)\ast p_{\rm obs}(iT_s+t),
\end{align}
where notation $\rm Pr(\cdot)$ and $\mathbb{E}(\cdot)$ are used to represent the probability function and the expectation operator, respectively.

For error probability analysis, we adopt the genie-aided decision feedback (DF) detector \cite{mosayebi2014receivers}. This detector benefits from the assistance of a genie, which informs it about the previously-transmitted bits, i.e., $\hat{B}_i = B_i$ for $i = 1,..., W$. If the correct values of the previously transmitted bits, $B_i = b_i$ for $i \in \{1, . . . , W\}$, are available to the decoder, and $\Pr(B_0 = 1) = \Pr(B_0 = 0) = \frac{1}{2}$, then the Maximum-A-Posteriori (MAP) detector for bit $B_0$, considering $\textbf{Y}_R = y$ received molecules in the current time slot, is formulated as
\begin{equation}\label{ML}
\hat{B}_0=\arg \max_{b_0\in \{0,1\}} \Pr(\textbf{Y}_R=y|b_0,b_1,...,b_W).
\end{equation}

Here, $\hat{B}_0$ represents the estimated transmitted bit in the current time slot. In practice, the previously-transmitted bits, $B_w = b_w$ for $j \in \{1,..., W\}$, are unknown. As a result, previous decisions, $\hat{B}_w = \hat{b}_w$ for $w\in \{1,..., W\}$, need to be utilized in \eqref{pois1} instead.

We can derive the threshold decision rule for receiver output in the current time slot, $y$, from \eqref{ML}. According to this rule, if $y \leq \rm \xi$, then $\hat{B_0}$ is set to $0$, and if $y > \xi$, then $\hat{B_0}$ becomes $1$. We calculate $\xi$ as follows \cite{schafer2020spherical}: 
\begin{equation}
\xi=\dfrac{y(t_s)}{\ln \left(1+\frac{y(t_s)}{I(t_s)}\right)}.
\end{equation}

To compute this threshold, having perfect knowledge of the channel, i.e., $p_{\rm obs}(t_s)$, is necessary. The error probability of this detector can be determined by the following expression:
\begin{equation}\label{pois2}
P_{\rm error}=\left(\frac{1}{2}\right)^{W+1}\sum_{b_0,b_1,..., b_W}\rm Pr(E|b_0,b_1,..., b_W),
\end{equation}
where $E$ represents an error event, and we have
\begin{align}\label{pois}
& \Pr(E|b_0,b_1,..., b_W)\\
&   =\sum_{y \overset{b_0=1}{\underset{b_0=0}{\lessgtr}} \xi}\dfrac{\mathrm{e}^{-\mathbb{E}((\textbf{Y}_R|b_0,b_1,...,b_W))}\left(\mathbb{E}((\textbf{Y}_R|b_0,b_1,...,b_W))\right)^y}{y!}.
\end{align}

\bibliographystyle{IEEEtran}
\bibliography{E2EREF2}

\end{document}